\documentclass[aps,prc,nofootinbib,showkeys showpacs,reprint,twocolumn]{revtex4-1}
\pdfoutput=1
\usepackage{graphicx}
\usepackage{dcolumn}
\usepackage{bm}
\usepackage{slashed}
\usepackage{amsmath,graphicx}
\usepackage[colorlinks=true,linktocpage=true,linkcolor=blue,citecolor=blue]{hyperref}
\usepackage{float}
\usepackage{nicefrac}
\usepackage[normalem]{ulem}
\usepackage{amsmath}
\usepackage{subfigure} 
\usepackage{bbold} 
\usepackage{placeins}
\usepackage[makeroom]{cancel}
\usepackage{stmaryrd} % for weird double brackets
\def\be{\begin{equation}}
	\def\ee{\end{equation}}
\newcommand{\bel}[1]{\begin{eqnarray}\label{#1}}
	\newcommand{\eel}{\end{eqnarray}}
\def\barr{\begin{array}}
	\def\earr{\end{array}}
\def\beq{\begin{eqnarray}}
	\def\eeq{\end{eqnarray}}
\def\bfig{\begin{figure}}
	\def\efig{\end{figure}}
\def\lt{\left}
\def\rt{\right}
\newcommand{\nn}{\nonumber}
\newcommand{\f}[2]{\frac{#1}{#2}}

\newcommand{\p}{\partial}

\def\b{\beta}

\newcommand{\sh}[1]{\sinh#1}
\newcommand{\ch}[1]{\cosh#1}

\newcommand{\lab}[1]{\label{#1}}
\def\nn{\nonumber}

\def\av{{\boldsymbol A}}
\def\bv{{\boldsymbol B}}

\begin{document}
 
    \title{Effect of thermal shear on longitudinal spin polarization in a thermal model} 
	\author{Wojciech Florkowski}
	\email{wojciech.florkowski@uj.edu.pl}
	 \affiliation{Institute of Theoretical Physics, Jagiellonian University,  PL-30-348 Krak\'ow, Poland}
	\author{Avdhesh Kumar} 
	\email{avdhesh.k@iitgn.ac.in} 
	\affiliation{Indian Institute of Technology Gandhinagar, Gandhinagar-382355, Gujarat, India}
	\author{Aleksas Mazeliauskas}
\email{aleksas.mazeliauskas@cern.ch}
\affiliation{Theoretical Physics Department, CERN, 1211 Geneva 23, Switzerland}
	\author{Radoslaw Ryblewski} 
	\email{radoslaw.ryblewski@ifj.edu.pl}
	\affiliation{Institute of Nuclear Physics Polish Academy of Sciences, PL-31-342 Krak\'ow, Poland}
	\date{\today} 
	\bigskip

\begin{abstract}
By including the recently introduced thermal shear term that contributes to the spin polarization vector at local equilibrium, we determine longitudinal polarization of $\Lambda$ hyperons emitted from a hot and rotating hadronic medium using the thermal model with single freeze-out. In our analysis, we consider the RHIC top energies and use the model parameters which were determined in the earlier analyses of particle spectra and elliptic flow. We confirm that, unlike the previous calculations done by using only the thermal vorticity, the thermal shear term alone leads to the correct sign of the quadrupole structure of the longitudinal component of the polarization three-vector measured in experiments. However, we find almost complete cancellation between thermal shear and vorticity terms, which eventually leads to disagreement with the data. To clarify the role played by velocity and temperature gradient terms, we present a systematic analysis of different contributions to the longitudinal polarization.
\end{abstract}
     
\date{\today}% It is always \today, today,

	
	
\maketitle
%
%\tableofcontents
% 
\section{Introduction}
\label{sec1:intro}
In non-central heavy-ion collisions at the relativistic beam energies, the produced medium carries a large orbital angular momentum transferred from the two colliding nuclei. A non-negligible part of such an initial orbital angular momentum can be subsequently transformed into the spin part -- an effect that can be revealed in the spin polarization of emitted particles~\cite{Liang:2004xn,Voloshin:2004ha,Voloshin:2017kqp,Liang:2004ph}. As a matter of fact, non-zero spin polarization of $\Lambda$ and $\bar{\Lambda}$ hyperons was measured by the STAR Collaboration at BNL~\cite{STAR:2017ckg,Adam:2018ivw}. Later on, the spin polarization of other particles ($K^*$, $\phi$) was also observed by the ALICE Collaboration at CERN~\cite{Acharya:2019vpe}. 

The results obtained by STAR indicate that the global spin polarization of $\Lambda$'s points along the direction perpendicular to the reaction plane, which resembles the magneto-mechanical Barnett effect~\cite{Barnett:1935} and the Einstein–de Haas effect~\cite{dehaas:1915}. This feature has been successfully explained by the relativistic hydrodynamic models~\cite{Becattini:2015ska,Karpenko:2016jyx,Xie:2017upb,Pang:2016igs,Becattini:2017gcx,Xie:2017upb}. In this approach, the basic quantity that governs the spin polarization effects is the thermal vorticity tensor~\cite{Becattini:2016gvu,Karpenko:2016jyx,Becattini:2017gcx} defined by the expression $\varpi_{\mu \nu}= -\frac{1}{2} (\partial_\mu \beta_\nu-\partial_\nu \beta_\mu)$, where the four-vector $\beta_\mu$ is defined as the ratio of the fluid flow vector $u_\mu$ and the local temperature $T$, i.e., $\beta_\mu = u_\mu/T$. The expression for thermal vorticity suggests that spin polarization effects are solely determined by the hydrodynamic fields $u^\mu(x)$ and $T(x)$ at freeze-out.

Due to the successes in explaining global polarization, it has been an intriguing experimental observation to subsequently find out that the predictions of hydrodynamic models~\cite{Becattini:2017gcx} fail to reproduce the momentum dependence of longitudinal spin polarization of $\Lambda$'s~\cite{Adam:2019srw} (i.e., the spin polarization along the beam ($z$) direction). For a recent review of this issue see~\cite{Becattini:2020ngo}. Hydrodynamic model calculations yield a quadrupole structure of momentum dependence of spin polarization along the beam axis, which has as an opposite sign compared to that found in the experiment~\cite{Niida:2018hfw}. This mismatch between theoretical and experimental results has been dubbed the sign problem~\cite{Becattini:2017gcx,Becattini:2020ngo,Weickgenannt:2020aaf,Speranza:2020ilk,Weickgenannt:2021cuo}. Recently, it has been proposed that the sign problem can be resolved \cite{Becattini:2021iol,Fu:2021pok} by including the previously overlooked shear-induced polarization governed by a symmetric tensor (thermal shear) defined by the formula $\xi_{\mu\nu}=\frac{1}{2} (\partial_\mu \beta_\nu+\partial_\nu \beta_\mu)$ ~\cite{Becattini:2021suc,Liu:2021uhn}. However, the agreement with the data is obtained only if, in addition, the temperature gradients in thermal vorticity and shear are neglected~\cite{Becattini:2021iol} or, the mass of the $\Lambda$ hyperon is replaced by the constituent mass of the strange quark~\cite{Fu:2021pok}.

This type of behavior asks for a clarification of the role played by various contributions included in $\varpi_{\mu \nu}$ and $\xi_{\mu\nu}$. To do so, in the present paper we consider the thermal model with single freeze-out (SF)~\cite{Broniowski:2001we}.  In the past, thermal models have been successfully used to describe various hadronic yields and spectra in the final stages of heavy-ion collisions~\cite{Cleymans:1992zc,Braun-Munzinger:2001hwo,Florkowski:2001fp,Becattini:2005xt,Andronic:2017pug}, therefore, it is natural to choose them to study the spin polarization of the emitted hadrons such as the $\Lambda$ and $\bar{\Lambda}$ hyperons. In our recent paper that used the single-freeze-out model~\cite{Florkowski:2019voj}, we have included only the contributions to spin polarization coming from the thermal vorticity. In this work, we take into account the additional effects due to the thermal shear.

We note that the single-freeze-out model has been used in the  past for Au+Au collisions at the top RHIC energies to describe various features of soft hadron production (particle yields, transverse-momentum spectra, elliptic flow, HBT radii). Thus, we can use the previous estimates of various model parameters to calculate the final spin polarization of particles at freeze-out. In particular, since the thermal vorticity $\varpi_{\mu \nu}$ and the thermal shear tensor $\xi_{\mu\nu}$ are the antisymmetric and symmetric combinations of the fluid field gradients, respectively, we can obtain them from the already known parametrizations of the hydrodynamic flow at freeze-out. Temperature gradients in the direction orthogonal to the freeze-out hypersurface are obtained in this case from the hydrodynamic equations, using the method developed in our previous paper~\cite{Florkowski:2019voj}.

The paper is organized as follows: In Sec. II we give a brief description of the thermal model with single freeze-out.  In Secs. III and IV we discuss the calculation of various components of the thermal vorticity $\varpi_{\mu \nu}$ and thermal shear $\xi_{\mu\nu}$. In order to gain more insight about the contributions from the temperature gradient terms, we split both the thermal vorticity and thermal shear tensors into two separate parts: $\varpi_{\mu \nu} = \varpi^{I}_{\mu \nu}+\varpi^{II}_{\mu \nu}$ and $\xi_{\mu\nu} = \xi^{I}_{\mu\nu}+\xi^{II}_{\mu\nu}$. Here the terms with superscript $I$ represent the contributions from velocity gradient terms, while the terms with superscript $II$ represent temperature gradient terms. In Sec. V we discuss various observables related to the spin polarization of particles. Finally, in Sec. VI we discuss our results.

\smallskip
{\it Notation and conventions:} 
For the Levi-Civita tensor $\epsilon^{\mu\nu\rho\sigma}$ we follow the convention $\epsilon^{0123} = -\epsilon_{0123} =+1$. The metric tensor is of the form $g^{\mu\nu}$ = diag$(+1, -1, -1, -1)$. We denote the scalar product of two four-vectors $A$ and $B$ as $A \cdot B =A^{\mu}B_{\mu}= g_{\mu \nu} A^\mu B^\nu = A^0 B^0 - \av \cdot \bv$, where bold font represents three-vectors.  The dual form of a rank-two antisymmetric tensor is represented by a tilde, 
\begin{align}
  \tilde{X}^{\mu\nu} = \frac{1}{2}\epsilon^{\mu\nu\alpha\beta}X_{\alpha\beta}.
 \label{equ:dualTensor}
\end{align}
Throughout the text we make use of natural units, $\hbar = c= k_B~=1$.

%%%%%%%%%%%%%%%%%%%%%%%%%%%%%%%%%%%%%%%%%%%%%%%%%%%%%%%%
%
\section{\label{sec:thermalmodel} Thermal model with single freeze out }
In its standard formulation, the thermal model with single freeze-out uses only four parameters: temperature $T$, baryon chemical potential $\mu_B$, proper time $\tau_f$, and system size $r_{\rm max}$.~\footnote{Other thermodynamic parameters, such as the strange and isospin chemical potentials, are not independent. They follow from the assumptions about the strangeness and charge neutrality of the system (valid for midrapidity region at relativistic energies). } The two thermodynamic parameters,  $T$ and $\mu_B$, are fitted from the ratios of hadronic abundances, while the two geometric ones, $\tau_f$ and $r_{\rm max}$, are obtained from the fits of experimental transverse-momentum spectra (they characterize the freeze-out hypersurface and the hydrodynamic flow). To be more specific, the freeze-out hypersurface is defined through the conditions: $\tau^2_f = t^2 - x^2 - y^2 - z^2$ and $x^2 + y^2 \leq r^2_{\rm max}$. The hydrodynamic flow is assumed to have a Hubble-like form $u^\mu = x^\mu/\tau$.

In this work, we use an extended version of the single-freeze-out model which may incorporate the phenomena related to the elliptic flow. This is an important aspect of the model, as the longitudinal polarization can be explained by a naive non-relativistic model that connects spin polarization with a rotation of the three-velocity field induced by the elliptic flow in the transverse plane~\cite{Voloshin} -- the sign problem appears only if one starts to use a relativistic description. 

In the extended version of the thermal model, we include the elliptic deformations of both: the emission region in the transverse plane and the transverse flow~\cite{Broniowski:2002wp}. The elliptic asymmetry in the transverse plane is included by the following parameterization of the boundary region
\beq 
x &=& r_{\rm max} \sqrt{1-\epsilon} \cos\phi, \nn \\
y &=& r_{\rm max} \sqrt{1+\epsilon} \sin\phi.
\eeq 
In the above equations, $\phi$ is the azimuthal angle, while $r_{\rm max}$ and $\epsilon$ are the model parameters. Note that here we take  $\epsilon >0$ which implies that the system formed in the collisions is elongated in the $y$ direction (out-of-plane). Accordingly, the asymmetric flow profile is given by the expression
\beq
u^\mu&=&\frac{1}{N}\left(t,~x\sqrt{1+\delta},~ y\sqrt{1-\delta },~z\right), \label{eq:u}
\eeq
where the parameter $\delta$ characterizes transverse flow anisotropy. The condition $\delta > 0$ corresponds to the case where there is more flow in the reaction plane (positive elliptic flow). The normalization factor can be obtained by using the normalization condition $u^{\mu}u_{\mu}=1$, which gives
\beq
N={\sqrt{\tau ^2-\left(x^2-y^2\right)\delta}}\,,
\label{eq:N}
\eeq
where the proper time $\tau$ is defined by 
\beq
\tau^2 = t^2 -x^2-y^2-z^2. 
\label{eq:tau}
\eeq
We use herein the values of $\epsilon$,   $\delta$, $\tau_f$, and $r_{\rm max}$ that have been used before to describe the PHENIX data for three different centrality classes at the beam energy $\sqrt{s_{NN}}=130$~GeV~\cite{baran,Florkowski:2004du}. They are collected in Table~\ref{tab}.

\begin{table*}
\centering
\begin{tabular}{ |p{3cm}||p{3cm}||p{3cm}||p{3cm}||p{3cm}|} 
   \hline
% \multicolumn{5}{|c|}{Thermal Model paramters used to describe PHENIX data  ($\sqrt{s_{NN}}=130$~GeV)} \\
%   \hline
 c $\%$ & $\epsilon$ &   $\delta$ &   $\tau_f$~[fm] &   $r_{\rm max}~[fm]$ \\
  \hline
   $0-15$ &   $0.055$ &   $0.12$ &   $7.666$ &   $6.540$ \\
  $15-30$ &   $0.097$ &   $0.26$ &   $6.258$ &   $5.417$ \\
  $30-60$ &   $0.137$ &   $0.37$ &   $4.266$ &   $3.779$ \\
 \hline
\end{tabular}
\caption{Geometric model parameters used to describe the PHENIX data at $\sqrt{s_{NN}}=130$~GeV for three different centrality classes \cite{baran,Florkowski:2004du}. The freeze-out temperature used in the calculation is $T_f =$~165~MeV. }
\label{tab}
\end{table*}

%%%%%%%%%%%%%%%%%%%%%%%%%%%%%%%%%%
\section{\label{sec:thvor} Components of thermal vorticity and thermal shear}

The thermal vorticity tensor is defined by the equation
\beq
\varpi_{\mu \nu} = -\frac{1}{2} (\p_\mu \b_\nu-\p_\nu \beta_\mu),
\label{eq:th_vor0}
\eeq
where $\beta_\mu=u^{\mu}/T$. It can be rewritten as a sum of the two terms
\beq
\varpi _{\mu \nu }=\underbrace{-\frac{1}{2T}\lt(\p_{\mu}u_{\nu}-\p_{\nu}u_{\mu}\rt)}_{\varpi^{I} _{\mu \nu }} + \underbrace{\frac{1}{2T^2}\lt( u_{\nu} \partial_\mu T-u_{\mu} \partial_\nu T \rt)}_{\varpi^{II} _{\mu \nu }}. \nonumber \\
\label{eq:th_vor1}
\eeq
Using the flow velocity defined by Eq.~(\ref{eq:u}), we can get the expressions for all the independent components of  ${\varpi^{I} _{\mu \nu }}$ as reported in Ref.~\cite{Florkowski:2019voj}. To calculate the temperature gradient term ${\varpi^{II} _{\mu \nu }}$ we use the hydrodynamic equations. Since we consider the top RHIC energies, we neglect the effects of baryon number density. We also expect that the effects of viscous terms are relatively small and consider perfect fluid equations of motion. In this case, we can use the equations discussed in appendix A of our previous paper \cite{Florkowski:2019voj}
\beq
\partial^{\alpha }T=T \left(D u^{\alpha }- c_s^2 u^{\alpha }\partial_\mu u^{\mu }\right),
\label{eq:tempgrad}
\eeq
where $c_s$ is the speed of sound taken to be $c_s=1/\sqrt{3}$. Using Eqs.~(\ref{eq:u})~and~(\ref{eq:tempgrad}) we can easily obtain the expressions for different components of the temperature gradient term ${\varpi^{II} _{\mu \nu }}$. We note that, they are the same as those obtained from ${\varpi^{I} _{\mu \nu }}$. 

The thermal shear tensor is defined by the formula
\beq
\xi_{\mu \nu} = \frac{1}{2} (\p_\mu \b_\nu+\p_\nu \beta_\mu),
\label{eq:th_shear0}
\eeq
which can further be written as a sum of two terms
\beq
\xi_{\mu \nu }=\underbrace{\frac{1}{2T}\lt(\p_{\mu}u_{\nu}+\p_{\nu}u_{\mu}\rt)}_{\xi^{I}_{\mu \nu }} \underbrace{- \frac{1}{2T^2}\lt(u_{\nu} \partial_\mu T +u_{\mu} \partial_\nu T \rt)}_{\xi^{II}_{\mu \nu }}. \nonumber \\
\label{eq:th_shear1}
\eeq
With the help of the flow velocity field defined  by Eq.~(\ref{eq:u}),  we  can get  the following expressions for all the independent components of  $\xi^{I}_{\mu \nu }$, namely
\beq
\xi^{I}_{01}&=&a \frac{t x}{2}  \left(1+\delta+\sqrt{1+\delta}\right), \nn \\
\xi^{I}_{02}&=&a \frac{t y}{2} \left(1-\delta+\sqrt{1-\delta }\right), \nn \\
\xi^{I}_{03}&=&a\, t z, \nn \\
\xi^{I}_{12}&=&-a \frac{x y \sqrt{1-\delta ^2}}{2}  \left(\sqrt{1+\delta}+\sqrt{1-\delta }\right), \nn \\
\xi^{I}_{23}&=-&a \frac{y z}{2} \left( 1-\delta+\sqrt{1-\delta } \right), \nn \\
\xi^{I}_{13}&=&- a \frac{x z}{2} \left(1+\delta+\sqrt{1+\delta}\right), \nn\\
\xi^{I}_{00}&=&-a\left(   (1+\delta) x^2+(1-\delta) y^2+z^2\right), \nn\\
\xi^{I}_{11}&=&a \sqrt{1+\delta } \left(-t^2+(1-\delta) y^2+z^2\right) , \nn\\
\xi^{I}_{22}&=&a\sqrt{1-\delta } \left(-t^2+(1+\delta) x^2+z^2\right), \nn\\
\xi^{I}_{33}&=&a\left((1+\delta) x^2+(1-\delta) y^2-t^2\right),
\label{eq:th_shear2}
\eeq
with $a=1/(TN^3)$. Similarly to the thermal vorticity case, we obtain the expressions for  different components of the temperature gradient term $\xi^{II}_{\mu\nu}$ (see Appendix~\ref{sec:xitempgrad}).  All the components of the thermal vorticity and thermal shear can be now calculated if we know five parameters: $T=T_f$ (freeze-out temperature), $r_{\rm max}$, $\tau_f$, $\epsilon$ and $\delta$. Values of these parameters can be taken from Refs.~\cite{baran,Florkowski:2004du}, see Table~\ref{tab}.

%%%%%%%%%%%%%%%%%%%%%%%%%%%%%%%%%%%%%%%%%%%%%%%%%%%%%%%%%%%%%%%
\section{\label{sec:level2} Spin polarization of particles at freeze-out}

\subsection{\label{subsec:level1} Pauli-Luba\'nski vector}

The spin polarization of particles with the four-momentum $p$ is characterized by the Pauli-Luba\'nski (PL) vector $\langle\pi^\star_{\mu}(p)\rangle$ calculated in their local rest frame where $p^\mu=p^{\mu\star}=(m,0,0,0)$. The average PL vector $\langle\pi_{\mu}(p)\rangle$ of particles with momentum $p$ is then given by the following ratio~\cite{Florkowski:2018ahw}
\beq
\langle\pi_{\mu}\rangle=\frac{E_p\frac{d\Pi _{\mu }(p)}{d^3 p}}{E_p\frac{d{\cal{N}}(p)}{d^3 p}}. \label{avPLV}
\eeq
Here $E_p\frac{d\Pi _{\mu }(p)}{d^3 p}$ is the total value of the PL vector of particles with momentum $p$, while ${E_p\frac{d{\cal{N}}(p)}{d^3 p}}$ is the invariant momentum density of all particles. The expressions for these two quantities are given below
\beq
E_p\frac{d\Pi _{\mu }(p)}{d^3 p} &=& -\f{ \cosh(\xi)}{(2 \pi )^3 m}
\int
e^{-\beta \cdot p} \,\Delta \Sigma \cdot p \,\,
\tilde{\omega }_{\mu \beta }p^{\beta }, \label{PDPLV}\\
E_p\frac{d{\cal{N}}(p)}{d^3 p}&=&
\f{4 \cosh(\xi)}{(2 \pi )^3}
\int
e^{-\beta \cdot p} \,\Delta \Sigma \cdot p  
\,. \label{eq:MD}
\eeq
Here $\Delta \Sigma _{\lambda}$ is an element of the freeze-out hypersurface. We note that we use here the classical Boltzmann statistics for $\Lambda$'s.

To carry out the integration over the freeze-out hypersurface,  we assume that the freeze-out takes place at a constant value of the proper time, i.e., at $\tau = \tau_f$. In this case, the three-dimensional element of the freeze-out hypersurface, $\Delta\Sigma_\lambda$, is given by the expression
\beq
\Delta \Sigma_{\lambda } &=& n_{\lambda }\, dx dy\,  d\eta \lab{sig}, 
\eeq
where
\beq
n^{\lambda } = \left(\sqrt{\tau^2_f+x^2+y^2}\ch\eta,x,y,
\sqrt{\tau^2_f + x^2+y^2}\sh\eta \right) \nn\\
\label{eq:n}
\eeq
with $\eta=\frac{1}{2} \ln \left[(t+z)/(t-z)\right]$ being the space-time rapidity. One can notice that \mbox{$n^\lambda n_\lambda = \tau^2_f$}.

We also use the parametrization of the particle four-momentum $p^\lambda$ in terms of the transverse momentum~$p_T = \sqrt{p_x^2+p_y^2}$  and  rapidity~$y_p$, 
\beq
p^\lambda &=& \left( E_p,p_x,p_y,p_z \right) = \left( m_T\ch y_p, p_x ,p_y, m_T\sh y_p \right), \nn\\ \lab{pl}
\eeq
where $m_T = \sqrt{m^2 + p_T^2}$ \, is the transverse mass and $m$ is the particle mass. For $\Lambda$'s we use the value $m=1.116$~GeV. Using Eqs.~(\ref{sig}) and (\ref{pl}) on the freeze-out surface, the scalar products $\Delta \Sigma \cdot p$  and $\beta \cdot p$ can be easily determined and the integration in Eq.~(\ref{eq:MD}) can be performed numerically.

To carry out the integration in Eq.~(\ref{PDPLV}), we assume that spin polarization tensor $\omega$ is expressed by the thermal vorticity and thermal shear. In this case, the contraction of the dual spin polarization tensor with the momentum four-vector can be written as
\begin{eqnarray}
\tilde{\omega }_{\mu \beta }p^{\beta }&=&\frac{1}{2} \epsilon _{\mu \beta \rho \sigma } p^{\beta } \left(\varpi ^{\rho \sigma }+2 \hat{t}^{\rho }\frac{p_{\lambda }}{E_p}\xi^{\lambda \sigma } \right),
\end{eqnarray}
where $\hat{t}^{\rho } = (1,0,0,0)$ \cite{Becattini:2021iol,Becattini:2021suc}.
Since we already know the components of $\varpi_{\mu\nu}$ and $\xi_{\mu\nu}$, we can easily determine the quantity $\tilde{\omega }_{\mu \beta }p^{\beta }$ and finally carry out the integrations in Eq.~(\ref{PDPLV}). In the next step we determine $\langle\pi_{\mu}(p)\rangle$ defined by Eq.~(\ref{avPLV}). As the experimental measurements are done in the central rapidity region, we consider the case of $y_p=0$ only. Furthermore, since we focus on the longitudinal spin polarization, we do not have to boost the four-vector $\langle\pi_{\mu}(p)\rangle$ to the particle rest frame, because $\langle\pi_{z}(p)\rangle$ is invariant under transverse boosts.

%%%%%%%%%%%%%%%%%%%%%%%%%%%%%%%%%%%%%%%%%%%%%%%%%%%%%%%%%%%%%
\subsection{\label{subsec:level2} Experimental observables describing spin polarization}

Equation \eqref{avPLV} can be used to calculate the contour plot of the longitudinal spin polarization in the transverse momentum plane $(p_x,p_y)$ at midrapidity, i.e., for $y_p =0$. However, it is also instructive to calculate the angular dependence of the longitudinal polarization for a fixed magnitude of the transverse momentum or to extract the angular dependence of the transverse momentum integrated longitudinal polarization. 

To mimic the experimental procedure~\cite{Adam:2019srw}, we need to weigh the spin polarization by the number of emitted particles first. Therefore, we multiply $\langle \pi_{z}(p)\rangle$ defined by Eq.~\eqref{avPLV} by the number of particles emitted into a particular phase-space region. Then, we can consider three convenient measures of the transverse-momentum dependence of the longitudinal polarization.

First, we consider the momentum integrated $n=2$ azimuthal harmonic of the longitudinal spin polarization at $y_p=0$,
\beq
\langle P_2\rangle&=&\frac{\frac{1}{2\pi}\int p_T dp_T d\phi_p \sin(2 \phi_p) \langle \pi_{z}(p)\rangle E_p\frac{{d \cal{N}}(p)}{d^3 p}}{\int p_Tdp_T d\phi_p E_p\frac{d{ \cal{N}}(p)}{d^3 p}}. \label{eq:azh2ini}
\eeq 
We note that $E_p\frac{d{ \cal{N}}(p)}{d^3 p} = \frac{d{ \cal{N}}(p)}{dy_p d^2 p_T}$ evaluated at $y_p=0$ counts the number of particles emitted per unit rapidity and cancels exactly the denominator in Eq.~\eqref{avPLV}. Therefore, Eq.~(\ref{eq:azh2ini}) simplifies to
\beq
\langle P_2\rangle
& = \frac{ \frac{1}{2\pi}\int d\phi_p p_T dp_T \sin(2 \phi_p) E_p\frac{d\Pi _{z }(p)}{d^3 p}}{\int d\phi_p p_T dp_T E_p\frac{d{ \cal{N}}(p)}{d^3 p}}\label{eq:azh2}
\eeq
with $E_p\frac{d\Pi _{z }(p)}{d^3 p}$ and $E_p\frac{d{ \cal{N}}(p)}{d^3 p}$ given by Eqs.~\eqref{PDPLV} and \eqref{eq:MD}. The second measure is useful if we want to calculate the angle dependence of the polarization averaged over some range of $p_T$. In this case we omit the azimuthal integration in the numerator in (\ref{eq:azh2}), but keep it in the normalization
\beq
\langle P(\phi_p)\rangle\equiv \frac{\int p_T dp_T E_p\frac{d\Pi _{z }(p)}{d^3 p}}{\int d\phi_p p_T dp_T E_p\frac{d{ \cal{N}}(p)}{d^3 p}}.  \label{eq:P_phi}
\eeq
%\vspace{0.1cm}
%
Finally, we can also look at the polarization anisotropy treated as a function of $p_T$. In this case, we perform azimuthal integrals in both the numerator and denominator, keeping the transverse momentum fixed, namely
\beq
\langle P(p_T)\rangle&=&\frac{ \frac{1}{2\pi}\int d\phi_p \sin(2 \phi_p) E_p\frac{d\Pi _{z}(p)}{d^3 p}}{\int d\phi_p E_p\frac{d{ \cal{N}}(p)}{d^3 p}}.\label{eq:P_pT}
\eeq
%

%%%%%%%%%%%%%%%%%%%%%%%%%%%%%%%%%%%%%%%
\section{Results and discussions}

\begin{table*}
\centering
\begin{tabular}{|p{1.5cm}||p{1.8cm}||p{1.8cm}||p{1.5cm}||p{1.7cm}||p{2cm}||p{2cm}||p{3cm}|} 
   \hline
% \multicolumn{5}{|c|}{Thermal Model paramters used to describe PHENIX data  ($\sqrt{s_{NN}}=130$~GeV)} \\
%   \hline
 c $\%$ & $\langle P_2\rangle_{\varpi^{I}}$ & $\langle P_2\rangle_{\varpi^{I}+\varpi^{II}}$  & $\langle P_2\rangle_{\xi^{I}}$ &   $\langle P_2\rangle_{\xi^{II}}$ &   $\langle P_2\rangle_{\xi^{I}+\xi^{II}}$ & $\langle P_2\rangle_{\varpi^{I}+\xi^{I}}$ &  $\langle P_2\rangle_{\varpi^{I}+\varpi^{II}+\xi^{I}+\xi^{II}}$ \\ 
  \hline 
   $0-15$ &   $-0.000026$ &   $-0.000052$ &   $0.000034$ &   $0.000018$ &  $0.000052$ &  $8.8\times10^{-6}$ &$-3.9\times10^{-7}$\\ 
     \hline
  $15-30$ &   $-0.000069$ &   $-0.000138$ &   $0.000076$ &   $0.000066$  & $0.000136$ &  $7.5\times10^{-6}$ & $-1.8\times10^{-6}$\\ 
    \hline
  $30-60$ &   $-0.000145$ &   $-0.000290$ &   $0.000156$ &   $0.000149$  &  $0.000277$ &  $0.0000104$ & $-0.0000134$\\ 
 \hline
\end{tabular}
\caption{Azimuthal harmonics due to various terms using the model parameters valid for different classes of collision centralities at $\sqrt{s_{NN}}=130$~GeV as shown in Table \ref{tab}. Temperature gradient contribution has been obtained using ideal equation of state ($c_s^2=1/3$). Integration of transverse momentum is carried out in the range %consistent with experiments (
$p_T$=0--3 GeV
%)
.}
\label{tab2}
\end{table*}
%
%\begin{table*}
%\centering
%\begin{tabular}{|p{1.5cm}||p{1.8cm}||p{1.8cm}||p{1.5cm}||p{1.7cm}||p{2cm}||p{2cm}||p{3cm}|} 
%   \hline
%% \multicolumn{5}{|c|}{Thermal Model paramters used to describe PHENIX data  ($\sqrt{s_{NN}}=130$~GeV)} \\
%%   \hline
% c $\%$ & $\langle P_2\rangle_{\varpi^{I}}$ & $\langle P_2\rangle_{\varpi^{I}+\varpi^{II}}$  & $\langle P_2\rangle_{\xi^{I}}$ &   $\langle P_2\rangle_{\xi^{II}}$ &   $\langle P_2\rangle_{\xi^{I}+\xi^{II}}$ & $\langle P_2\rangle_{\varpi^{I}+\xi^{I}}$ &  $\langle P_2\rangle_{\varpi^{I}+\varpi^{II}+\xi^{I}+\xi^{II}}$ \\ 
%  \hline 
%   $0-15$ &   $-0.000032$ &   $-0.000064$ &   $0.000044$ &   $0.000022$ &  $0.000066$ &  $0.000012$ &$2.0\times10^{-6}$\\ 
%     \hline
%  $15-30$ &   $-0.000085$ &   $-0.000171$ &   $0.000099$ &   $0.000081$  & $0.000173$ &  $0.000013$ & $-1.9\times10^{-6}$\\ 
%    \hline
%  $30-60$ &   $-0.000180$ &   $-0.000360$ &   $0.000203$ &   $0.000184$  &  $0.000353$ &  $0.000021$ & $-7.6\times10^{-6}$\\ 
% \hline
%\end{tabular}
%\caption{Azimuthal harmonics due to various terms using the model parameters used for different classs of collisions centralities at ($\sqrt{s_{NN}}=130$~GeV) as shown in \ref{tab}. Temperature gradient contribution has been obtained using ideal EoS ($c_s^2=1/3$). Integration of transverse mometum is carried out in the range consistent with experiments ($p_T$=0.5--6 GeV).}
%\label{tab3}
%\end{table*}
In this section, we present our numerical results describing the transverse-momentum dependence of the longitudinal spin polarization of the $\Lambda$ hyperons. In order to obtain numerical results we consider the expressions for total ${\varpi _{\mu \nu }}$ and $\xi_{\mu \nu }$ in the linear order in $\delta$. For each centrality class, we present two-dimensional plots showing the spin polarization dependence on $p_x$ and $p_y$, give the second azimuthal harmonic of the longitudinal spin polarization $\langle P_2\rangle$, show the azimuthal angle ($\phi_p$) dependence of the longitudinal spin polarization $\langle P(\phi_p)\rangle$ and the transverse momentum dependence of the second azimuthal harmonic of longitudinal spin polarization $\langle P(p_T)\rangle$. 

We use the single freeze-out model parameters as given in Table~\ref{tab}.
The values of the parameters were previously used to describe the PHENIX data at the beam energy $\sqrt{s_{NN}}=130$~GeV~\cite{baran}, for the three different centrality classes: $c$=0--15\%, $c$=15--30\%, and $c$=30--60\%,  at freeze-out temperature $T_f=0.165$~GeV. We do the $p_T$ integral in the range from 0 to 3 GeV and we checked that using the experimental momentum range 0.5-6 GeV does not change the results significantly.

\begin{figure*}
      \centering
		\subfigure[]{}\includegraphics[width=0.32\textwidth]{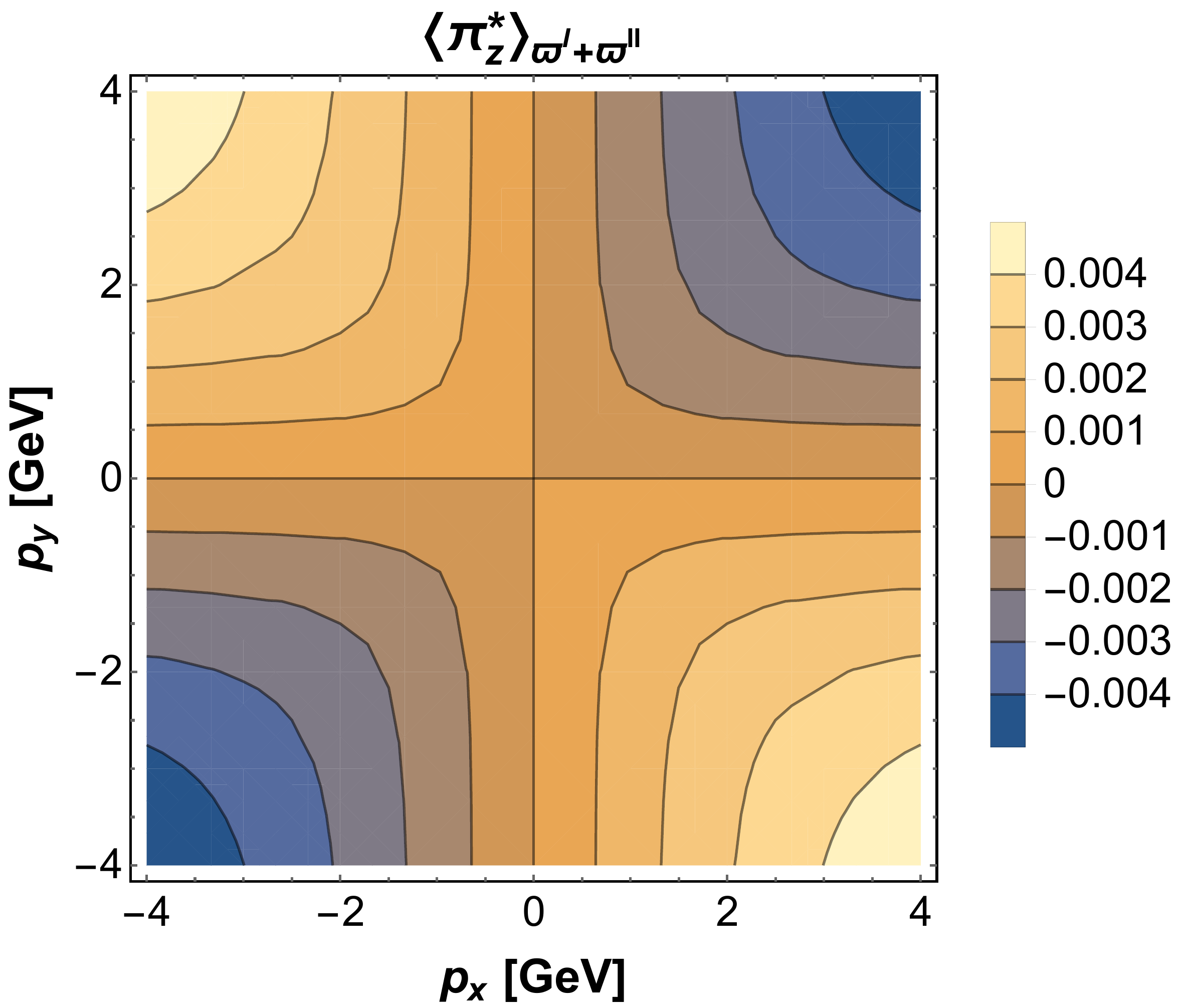}
		\subfigure[]{}\includegraphics[width=0.32\textwidth]{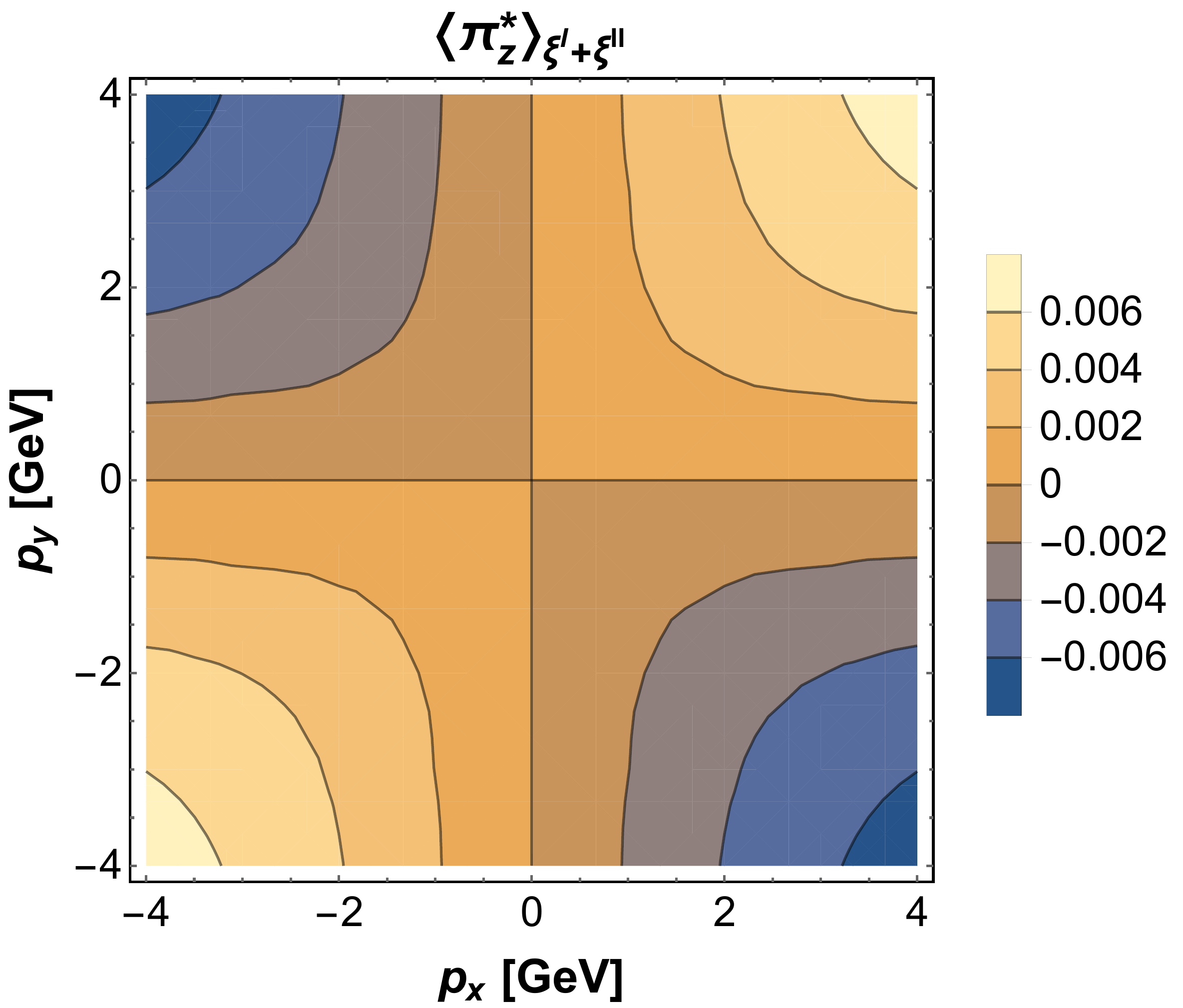}\\
		\subfigure[]{}\includegraphics[width=0.32\textwidth]{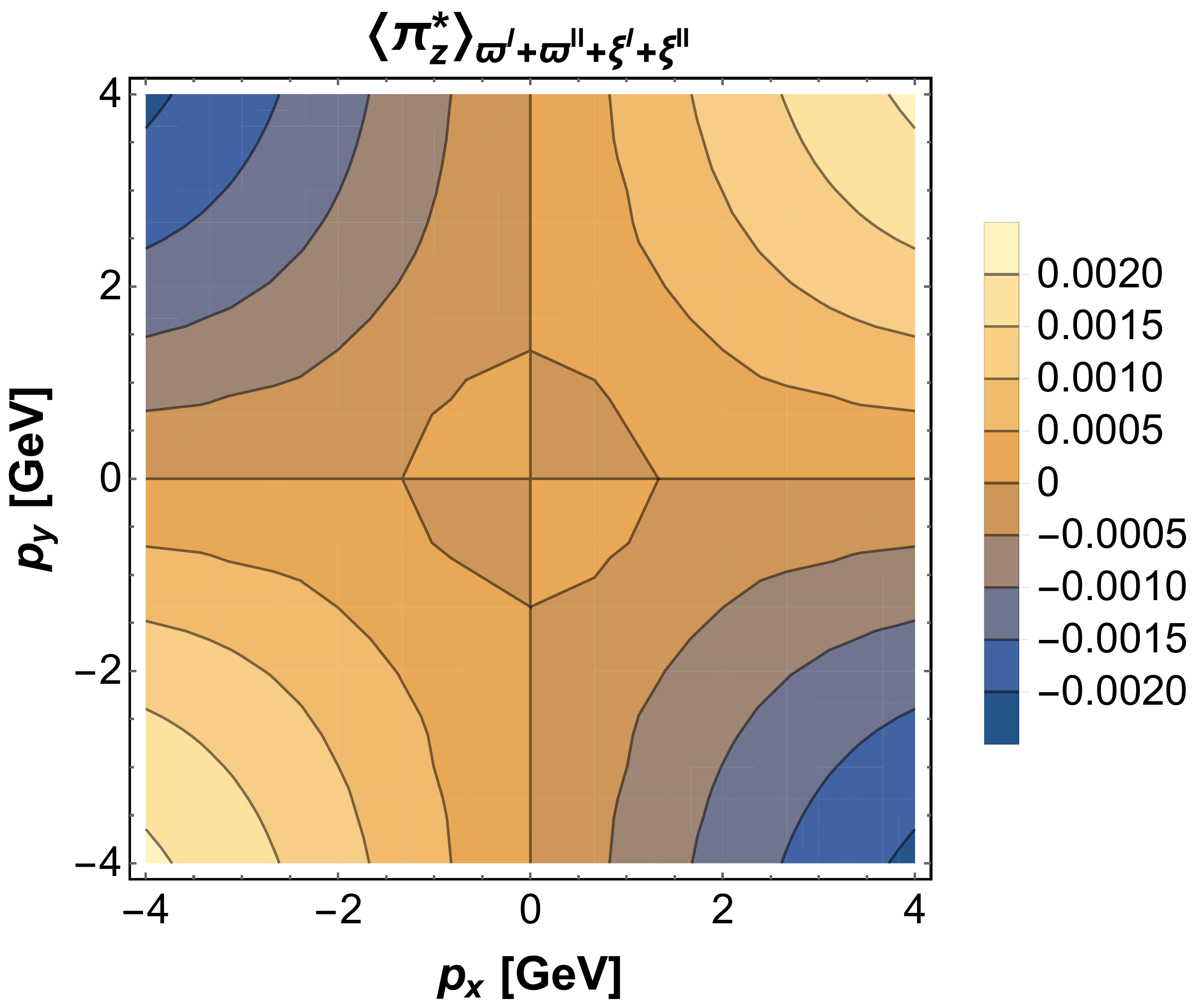}
		\subfigure[]{}\includegraphics[width=0.32\textwidth]{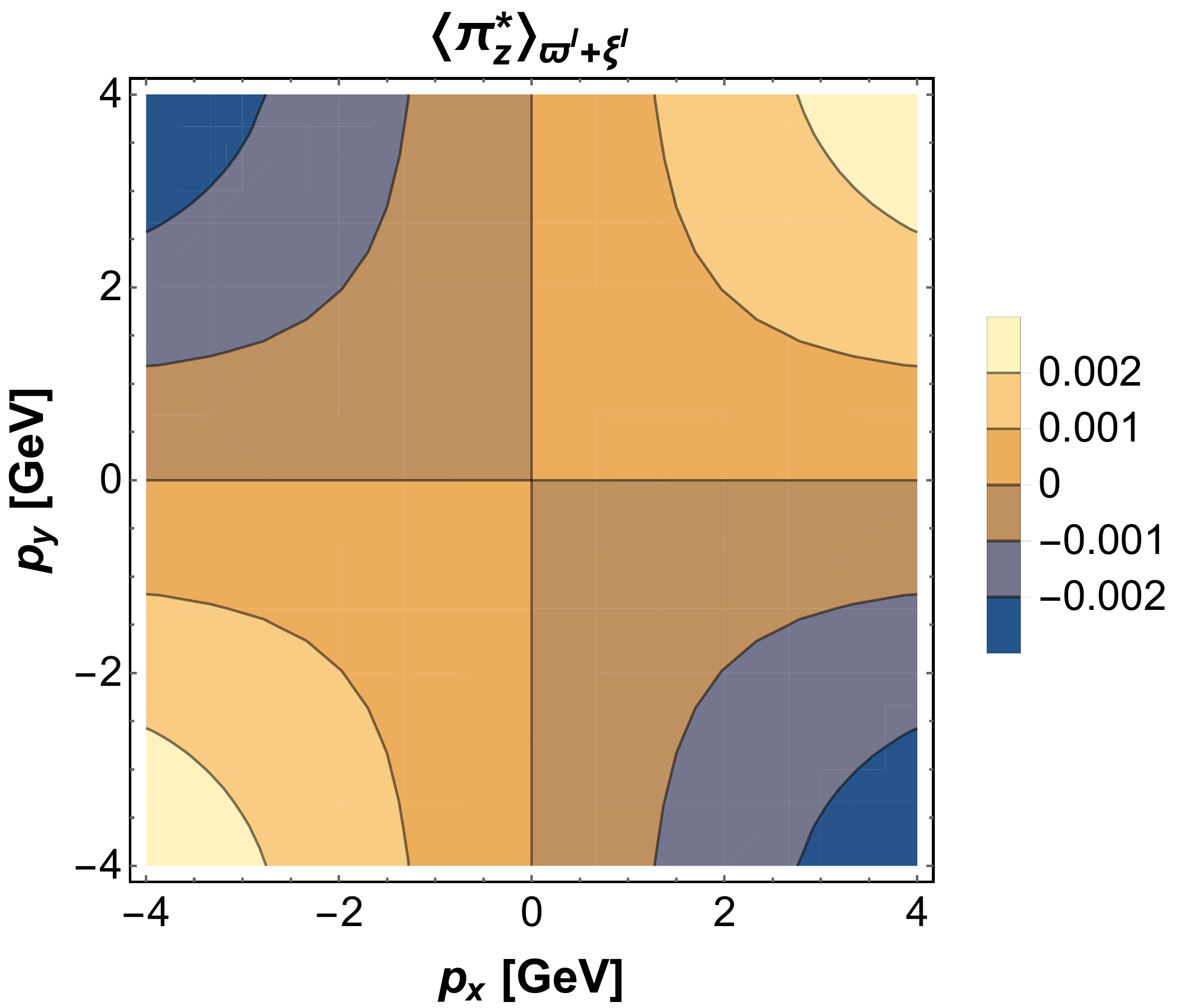}
		\caption{The longitudinal component of the mean spin polarization three-vector of $\Lambda$ hyperon as a function of its transverse momentum for the centrality class $c$=0--15\% and for different contributions defined in Sec.~\ref{sec:thvor}.
		%with temperature gradient terms evaluated using $c_s^2=1/3$. 
	 (a) the spin polarization  given by total thermal vorticity $\varpi^I+\varpi^{II}$, 
		 (b) the contribution from total thermal shear $\xi^I+\xi^{II}$,  (c)  the spin polarization given by all the terms in thermal vorticity and thermal shear, (d)  the contribution from the first terms of thermal vorticity and thermal shear tensor $\varpi^I+\xi^{I}$ (without temperature gradients contribution).
		} 
	 \label{fig:polarization1}
\end{figure*}

Figure~\ref{fig:polarization1} shows the two-dimensional plot of the spin polarization dependence on $p_x$ and $p_y$ for the centrality class $c$=0--15\%.
From panels (a) and (b) of Fig.~\ref{fig:polarization1} we see that thermal vorticity and thermal shear lead to opposite quadrupole structure of the longitudinal $\Lambda$ hyperon spin polarization. When both terms are added, as is done in panel (c),  the quadrupole structure of the total polarization changes sign as a function of $p_T$. At larger momentum, we recover the polarization structure seen in the experiment. In panel (d) we also show the results of neglecting temperature gradients as was done in Ref.~\cite{Becattini:2021iol}. 

To quantitatively characterize the quadrupole structure of the longitudinal spin polarization, in Table~\ref{tab2}  we present the second azimuthal harmonic  $\langle P_2 \rangle$ determined for three different centrality classes. We observe that the total thermal vorticity and thermal shear contributions to the $n=2$ harmonic of the longitudinal spin polarization are close in magnitude but have opposite signs. The sum of the two gives a very small negative polarization value. Neglecting temperature gradients does change the sign of the net polarization, but the value remains an order of magnitude smaller than individual thermal vorticity or thermal shear contributions.

\begin{figure*}
      \centering
		\subfigure[]{}\includegraphics[width=0.50\textwidth]{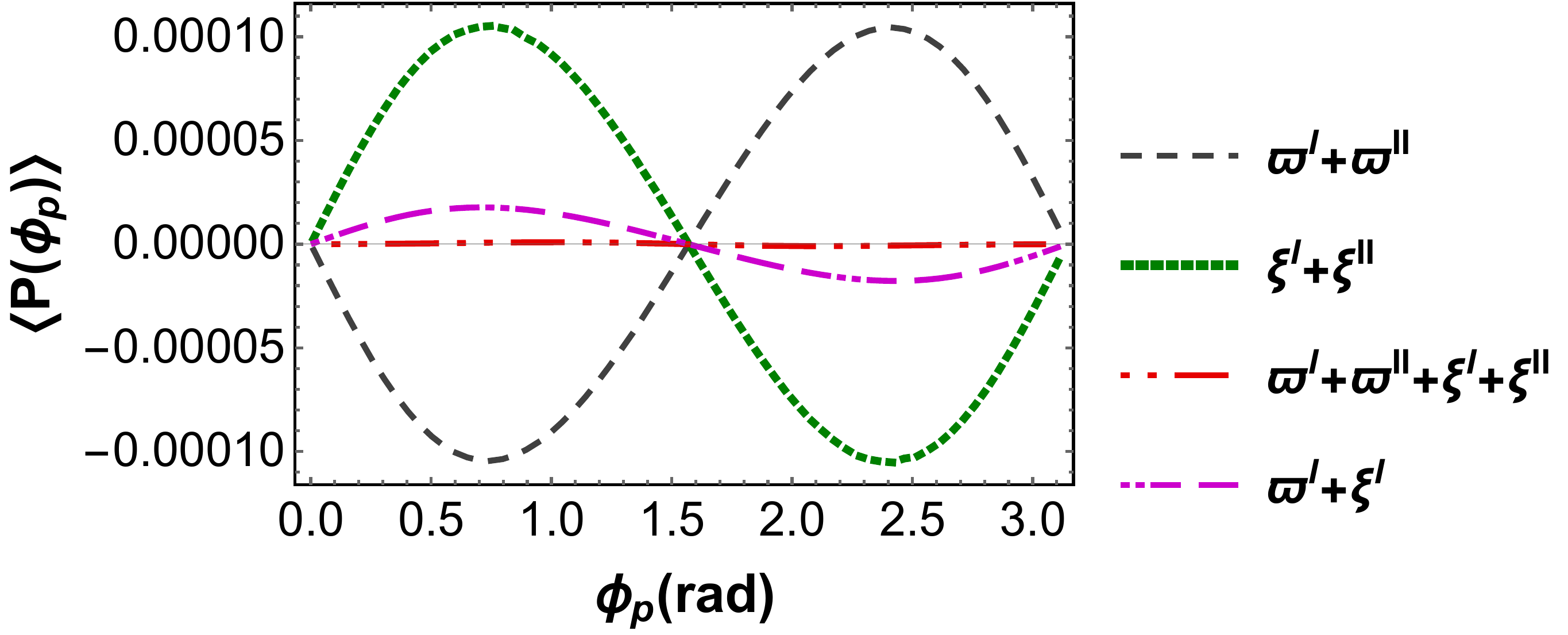}
			\subfigure[]{}\includegraphics[width=0.36\textwidth]{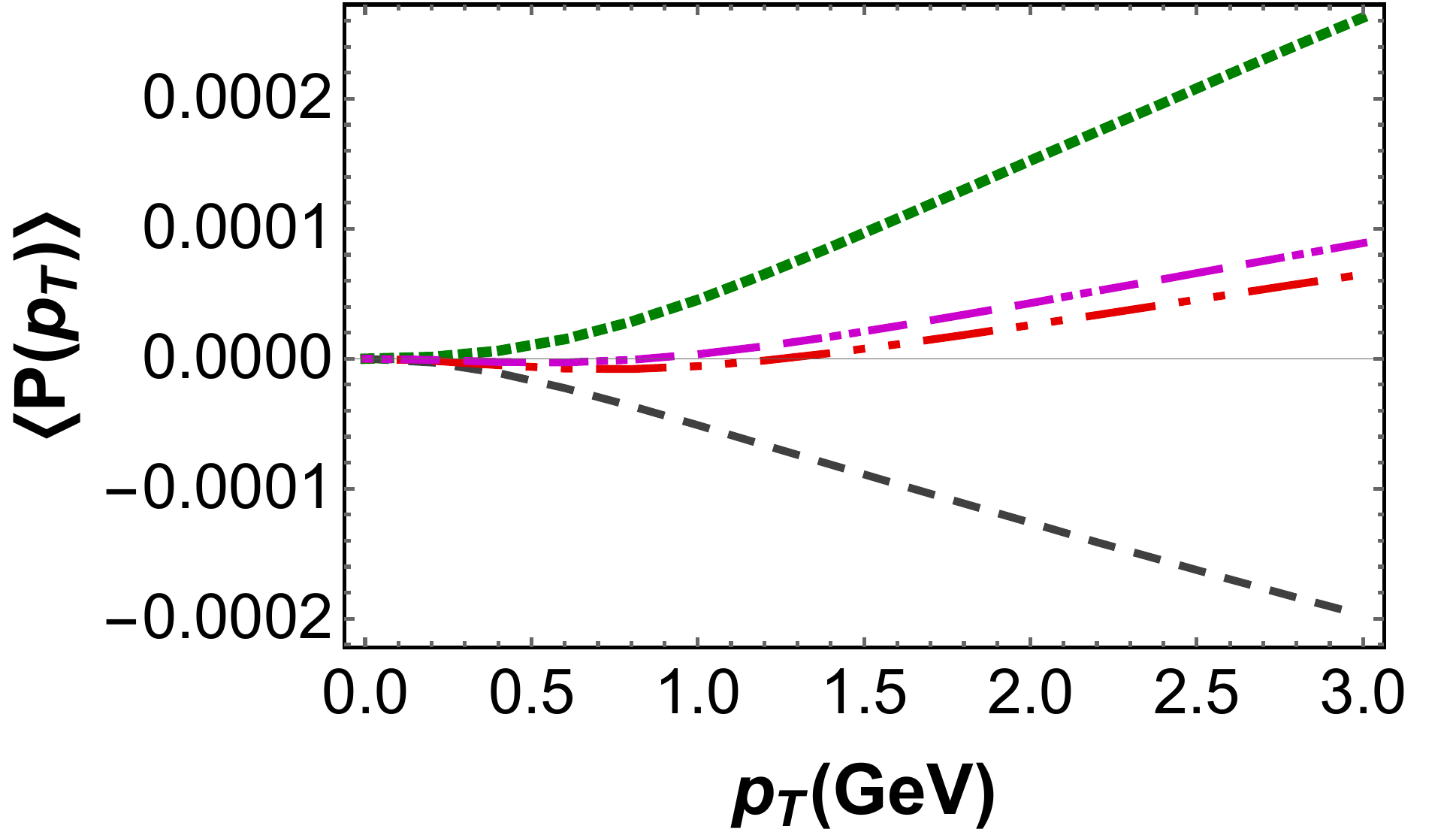}
		\caption{				(a) Azimuthal angle dependence of $p_T$-integrated (range $p_T$=0--3 GeV) longitudinal spin polarization \eqref{eq:P_phi} for the centrality class $c$=0--15\%. Lines correspond to different combinations of thermal vorticity and thermal shear tensor components defined in Sec.~\ref{sec:thvor}.
		(b) Transverse-momentum dependence of $n=2$ harmonic of longitudinal spin polarization. \eqref{eq:P_pT}
		}
		 \label{fig:f2_1new}
\end{figure*}

In Fig.~\ref{fig:f2_1new} (a) we show the azimuthal-angle dependence of the $p_T$-integrated longitudinal spin polarization \eqref{eq:P_phi}. The black dashed and green dotted lines show thermal vorticity and thermal shear contributions, respectively. The red and purple dot-dashed curves show the result of the net polarization with and without temperature gradients. With temperature gradients included, the net results are practically zero while dropping them gives the polarization dependence of the same sign as that of the thermal shear contribution.

In Fig.~\ref{fig:f2_1new} (b) we show the $n=2$ azimuthal harmonic of the longitudinal spin polarization treated as a function of transverse momentum. This clearly illustrates that thermal vorticity and thermal shear dominate in different momentum ranges. At low momenta, the thermal vorticity contribution is larger, but for $p_T>1\,\text{GeV}$ the thermal shear contribution is dominant. Since the total polarization is found by weighing the differential distribution with particle spectra, the negative low momentum polarization nearly exactly cancels the high-momentum positive polarization. By dropping the temperature gradient terms, one shifts the balance in favor of a positive azimuthal harmonic.

\begin{figure*}
     \centering
		\subfigure[]{}\includegraphics[width=0.32\textwidth]{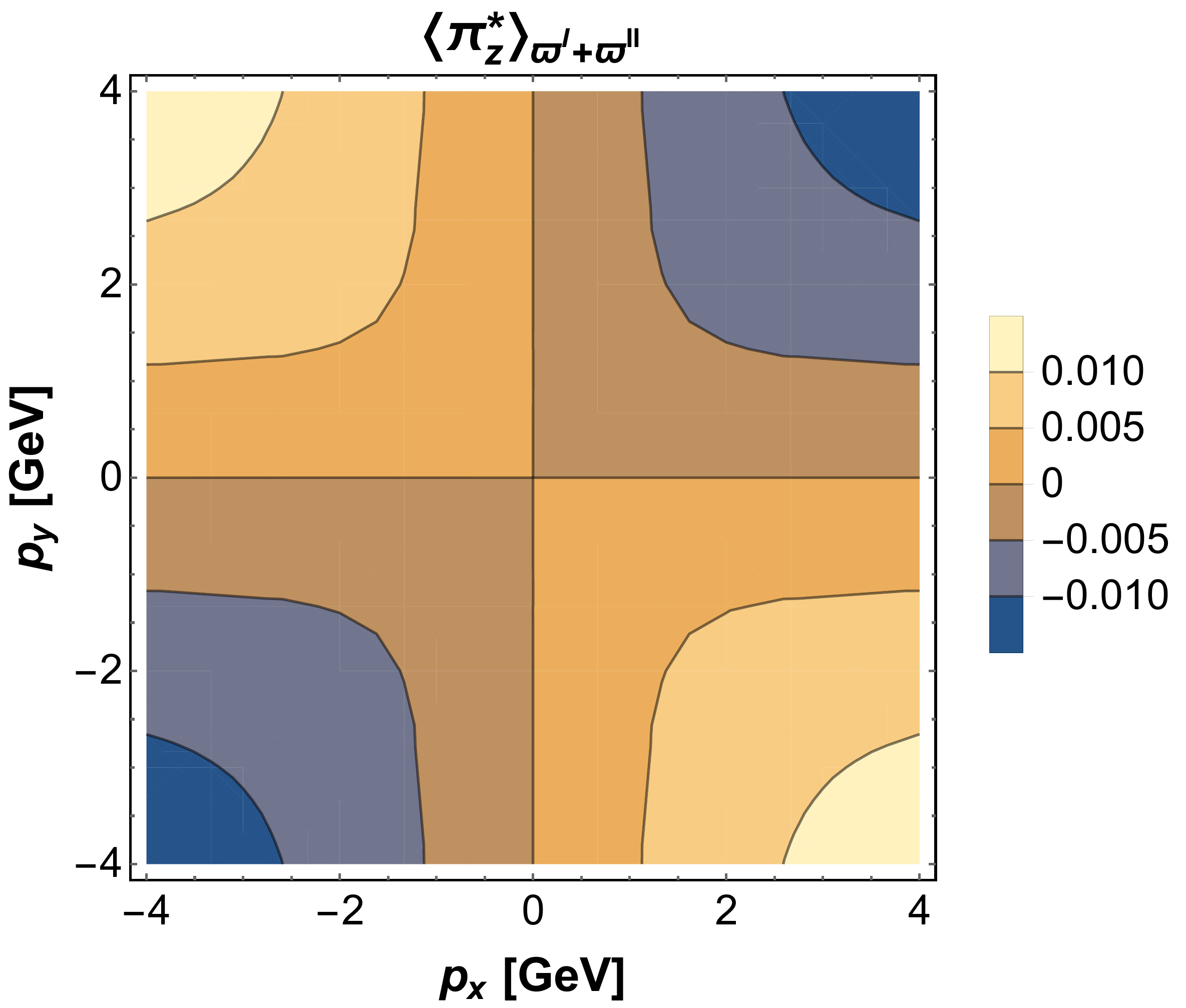}
		\subfigure[]{}\includegraphics[width=0.32\textwidth]{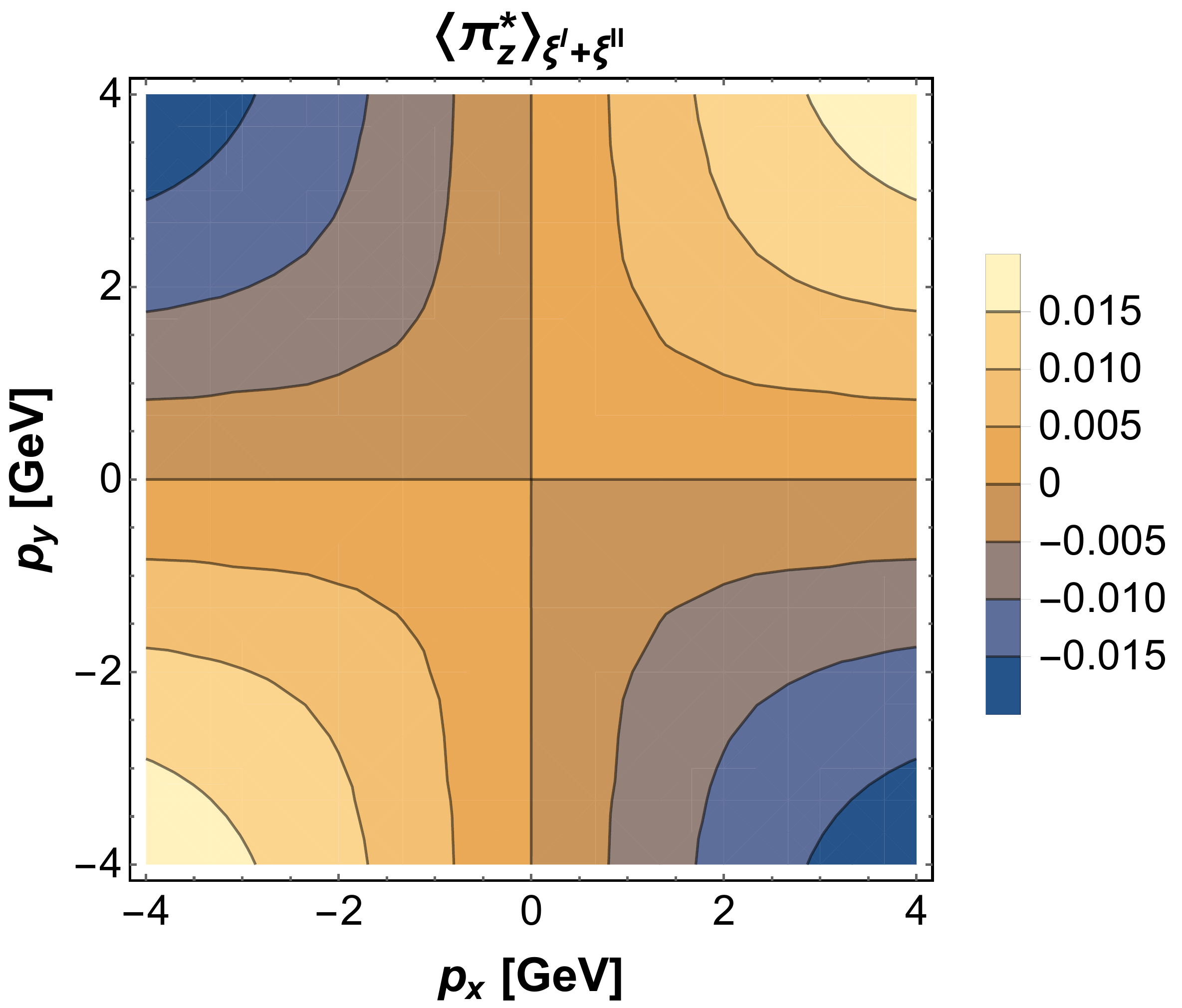}\\
		\subfigure[]{}\includegraphics[width=0.32\textwidth]{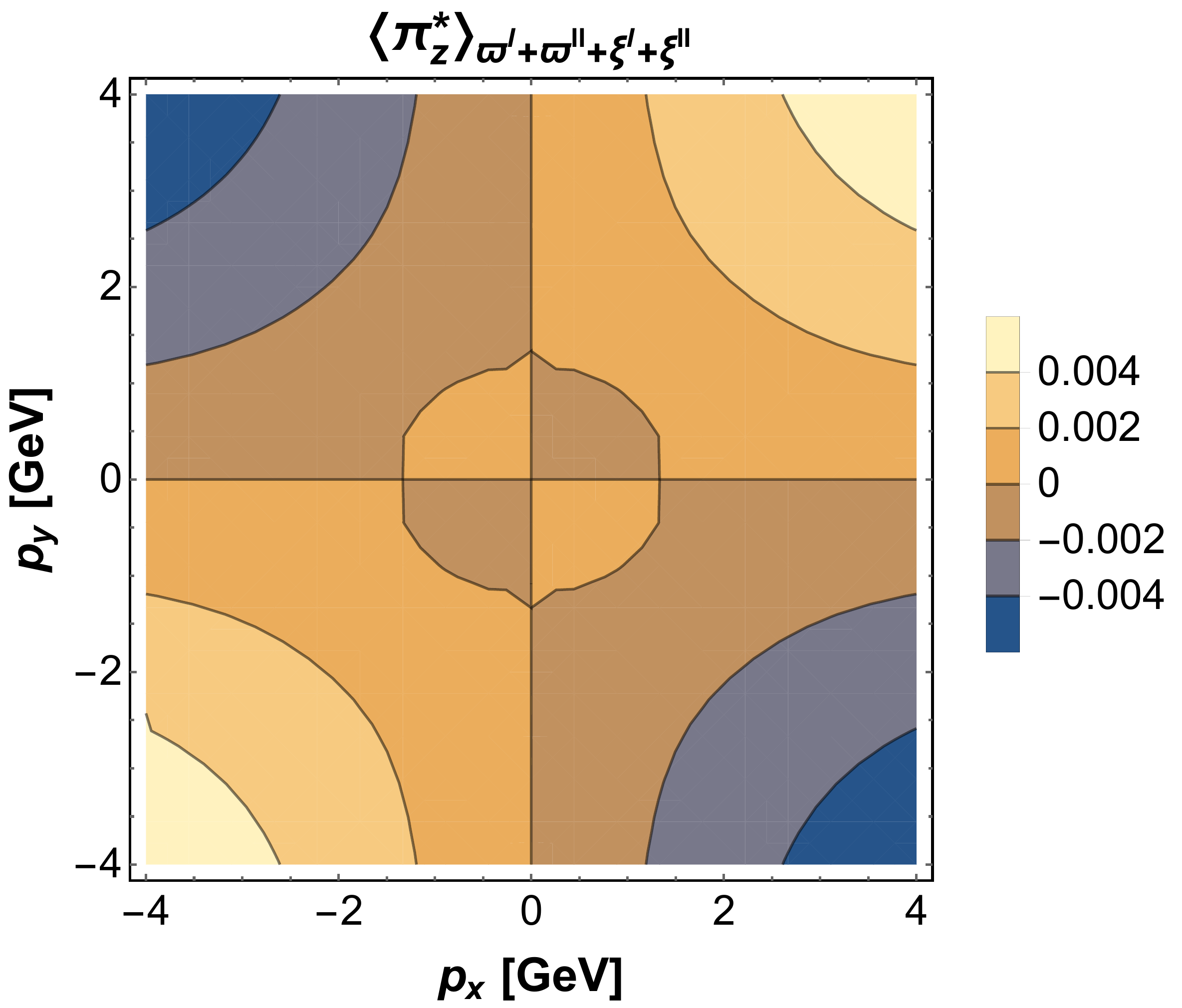}
		\subfigure[]{}\includegraphics[width=0.32\textwidth]{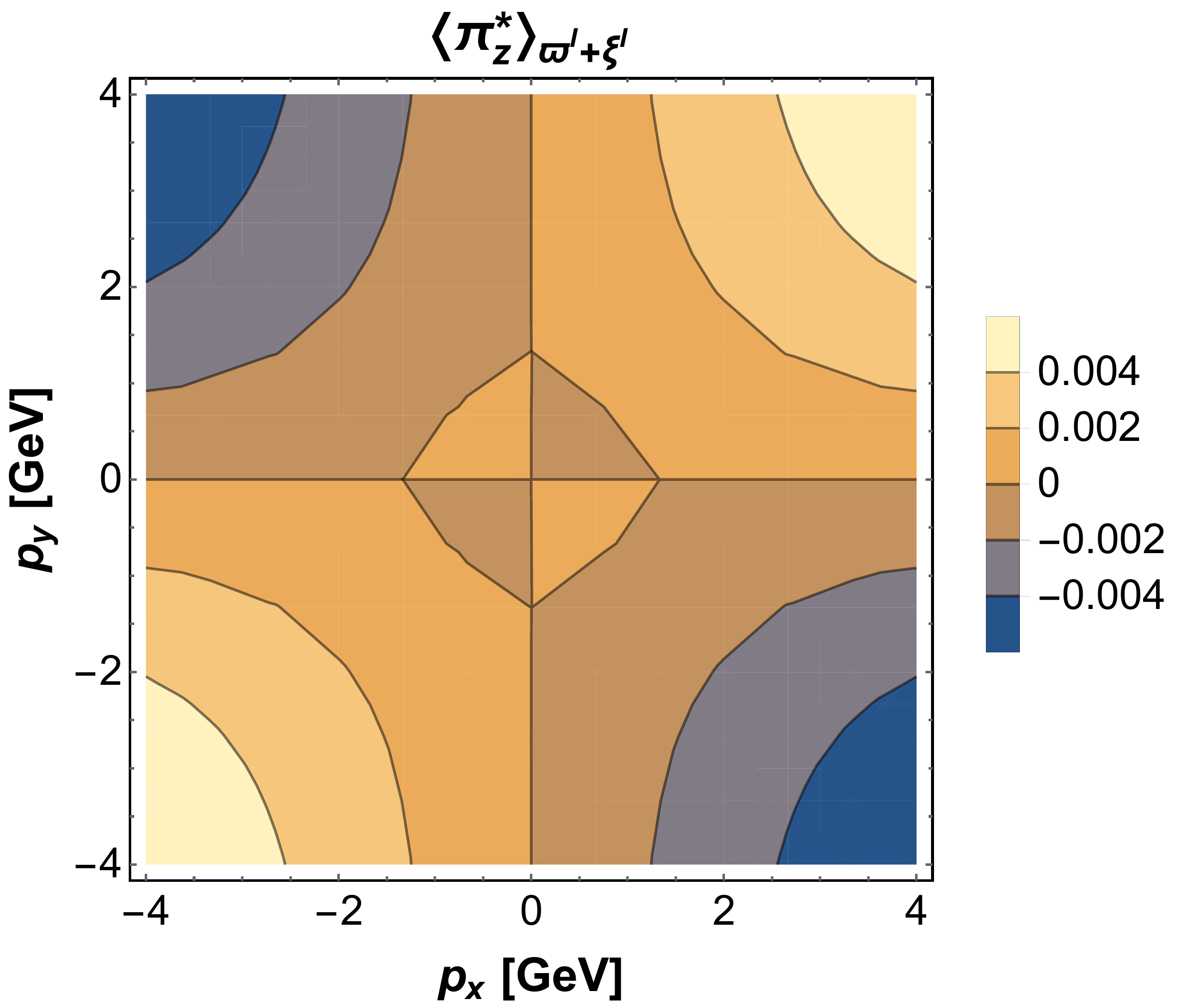}
		\caption{Same as Fig. \ref{fig:polarization1} but for the centrality class  $c$=15--30\%.} 
	 \label{fig:polarization2}
\end{figure*}

\begin{figure*}
      \centering
		\subfigure[]{}\includegraphics[width=0.50\textwidth]{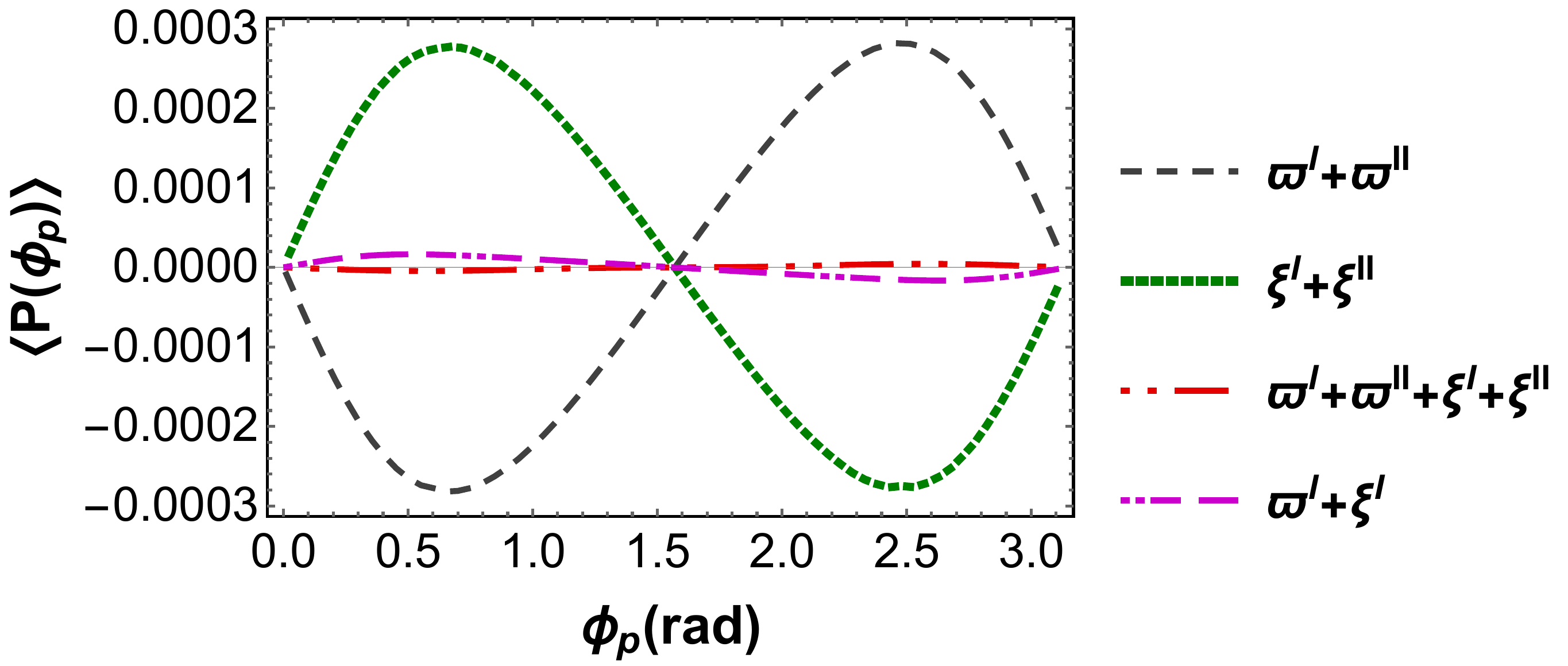}
		\subfigure[]{}\includegraphics[width=0.36\textwidth]{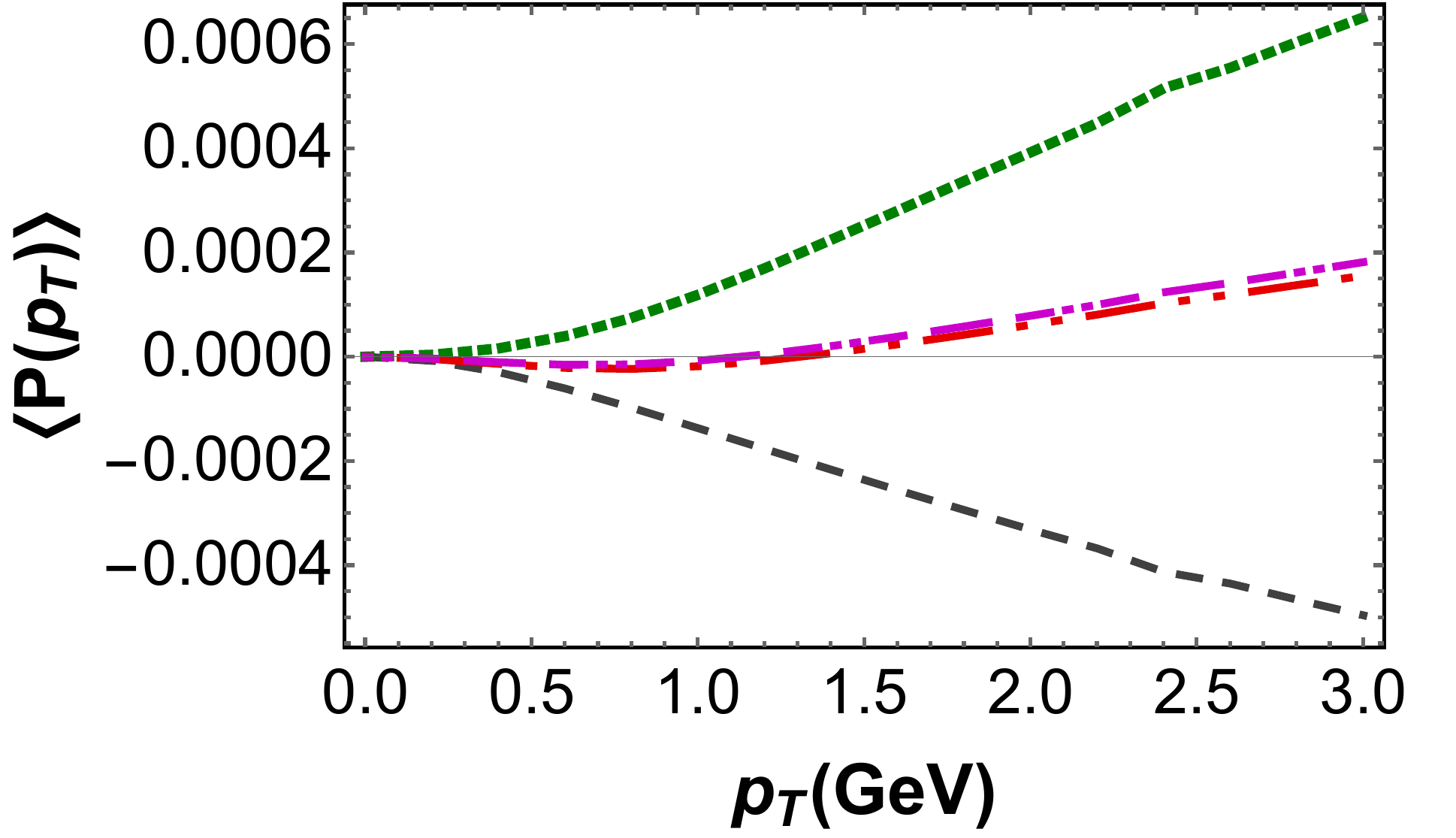}
		\caption{Same as Fig. \ref{fig:f2_1new} but for the centrality class  $c$=15--30\%.
		} 
	 \label{fig:f2_2new}
\end{figure*}

\begin{figure*}[ht!]
     \centering
		\subfigure[]{}\includegraphics[width=0.32\textwidth]{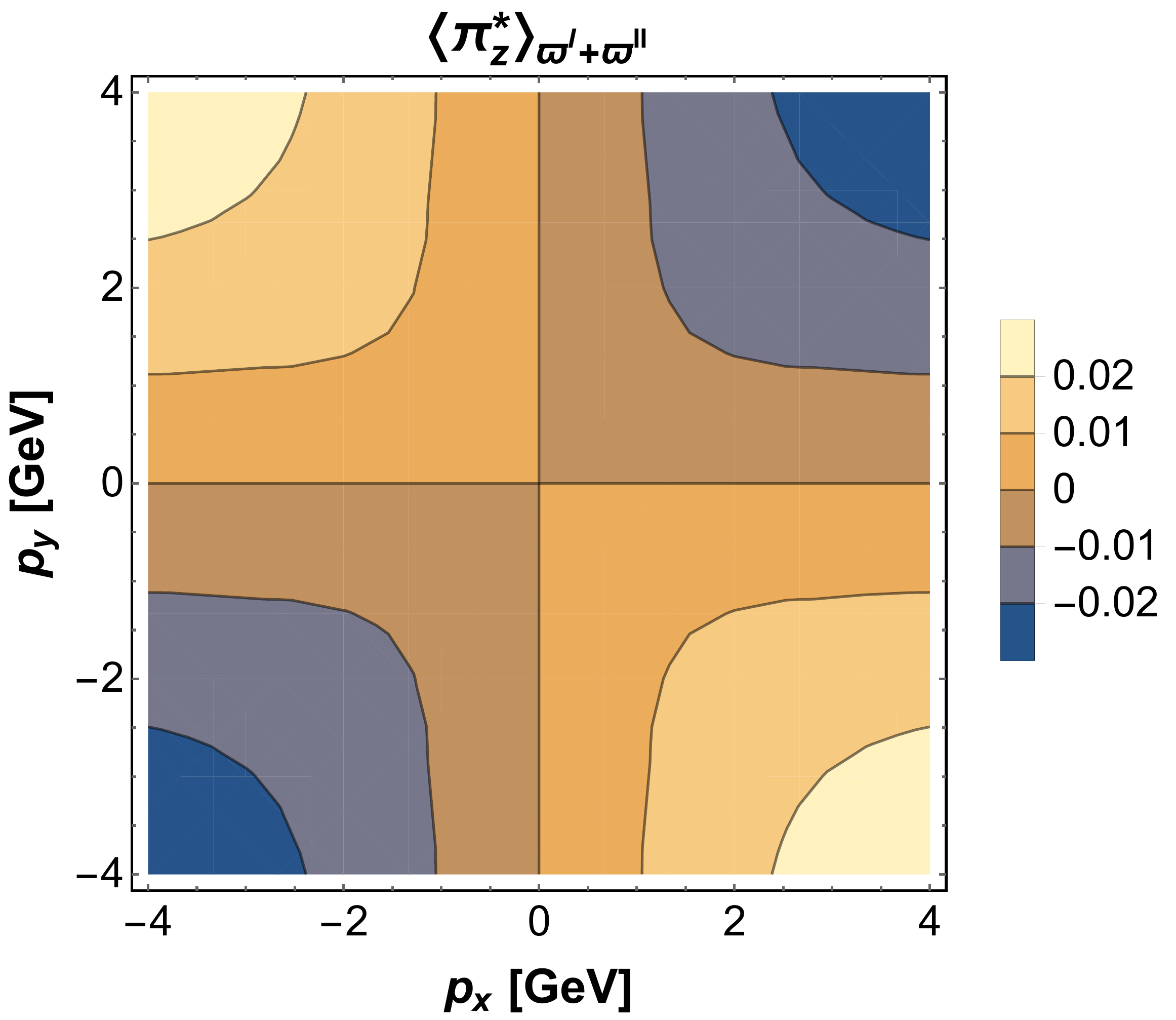}
		\subfigure[]{}\includegraphics[width=0.32\textwidth]{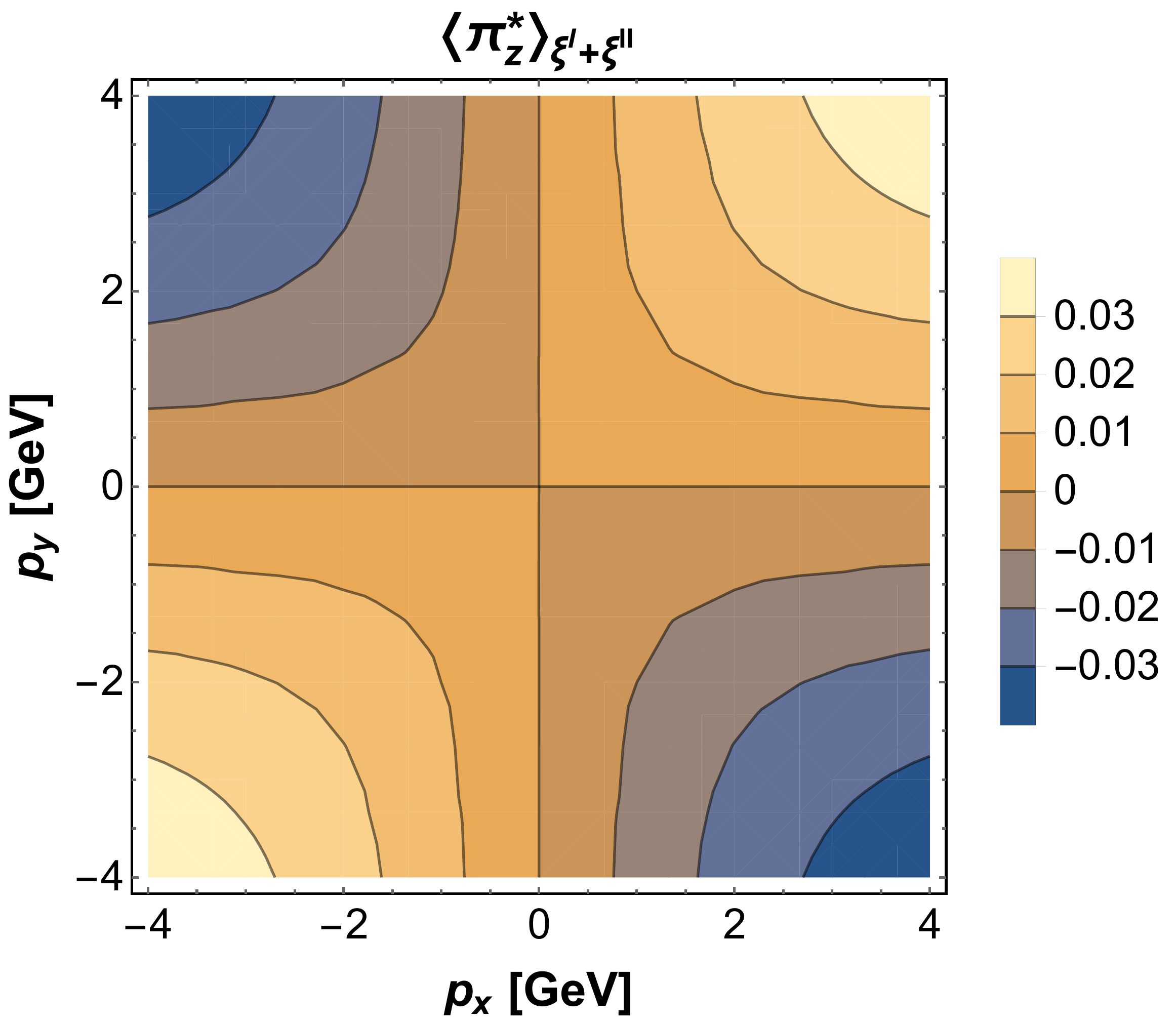}\\
		\subfigure[]{}\includegraphics[width=0.32\textwidth]{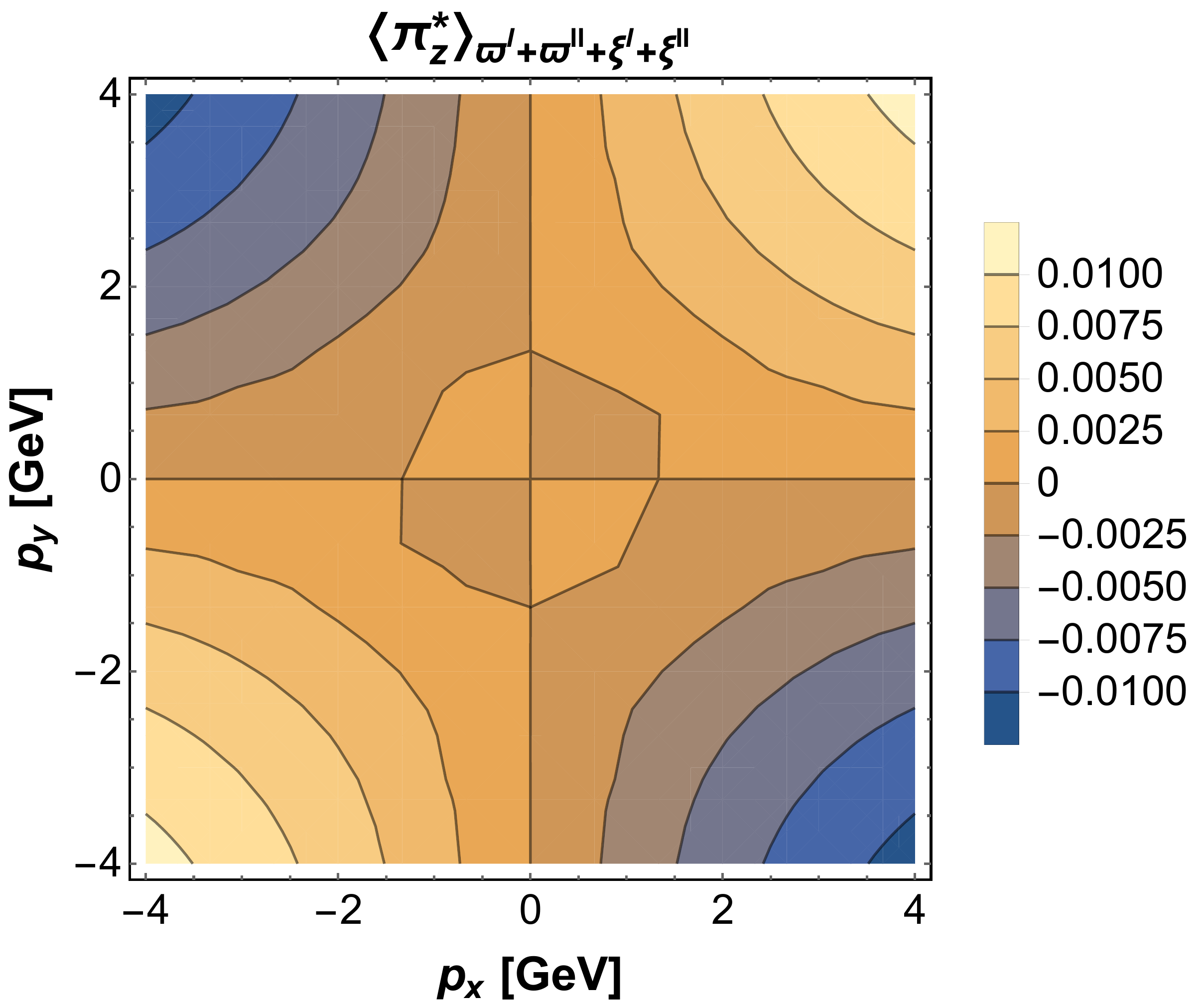}
		\subfigure[]{}\includegraphics[width=0.32\textwidth]{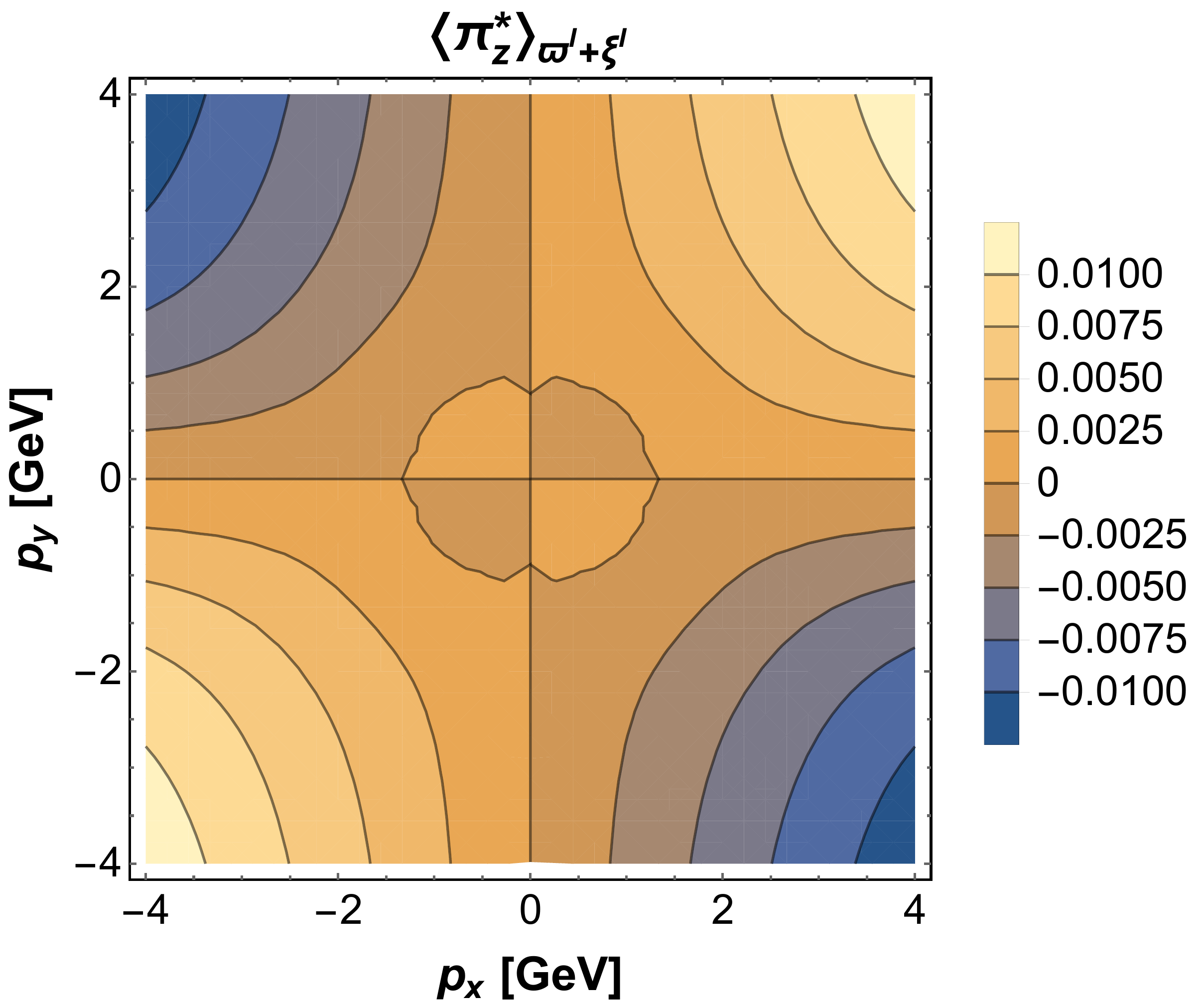}
		\caption{Same as Fig. \ref{fig:polarization1} but for the centrality class $c$=30--60\%.} 
	 \label{fig:polarization3}
\end{figure*}

\begin{figure*}
   \centering
		\subfigure[]{}\includegraphics[width=0.55\textwidth]{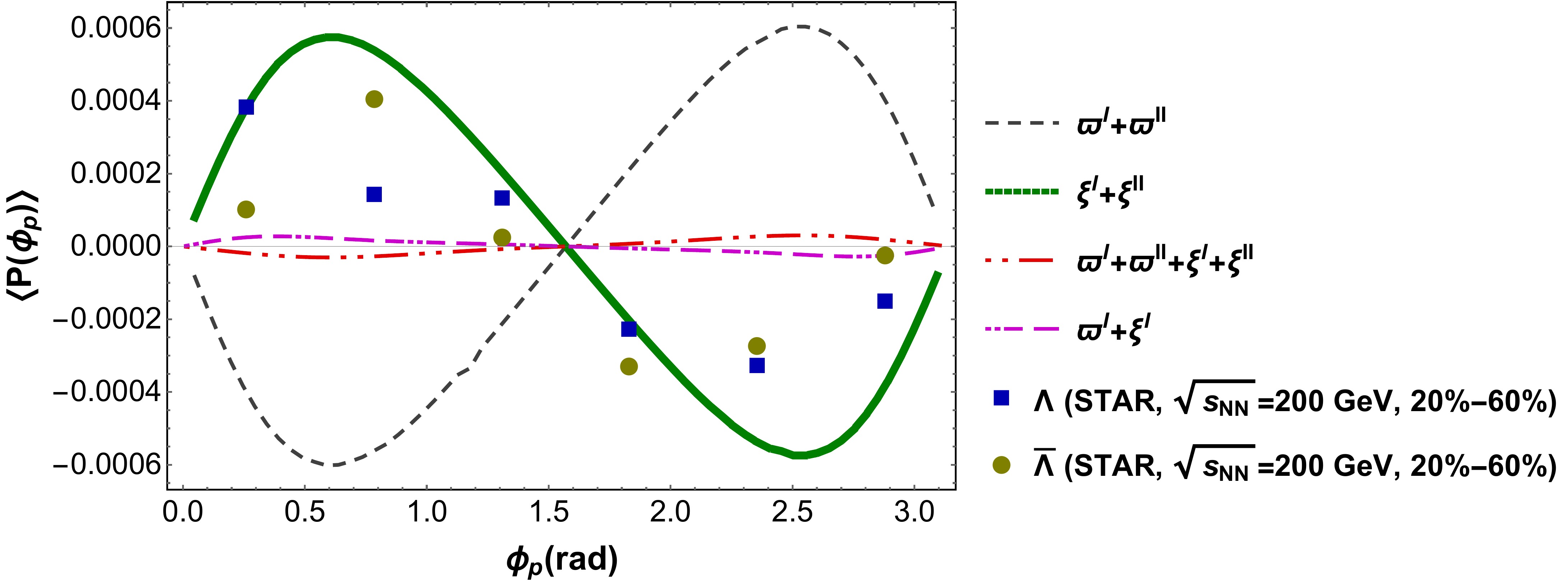}
		\subfigure[]{}\includegraphics[width=0.36\textwidth]{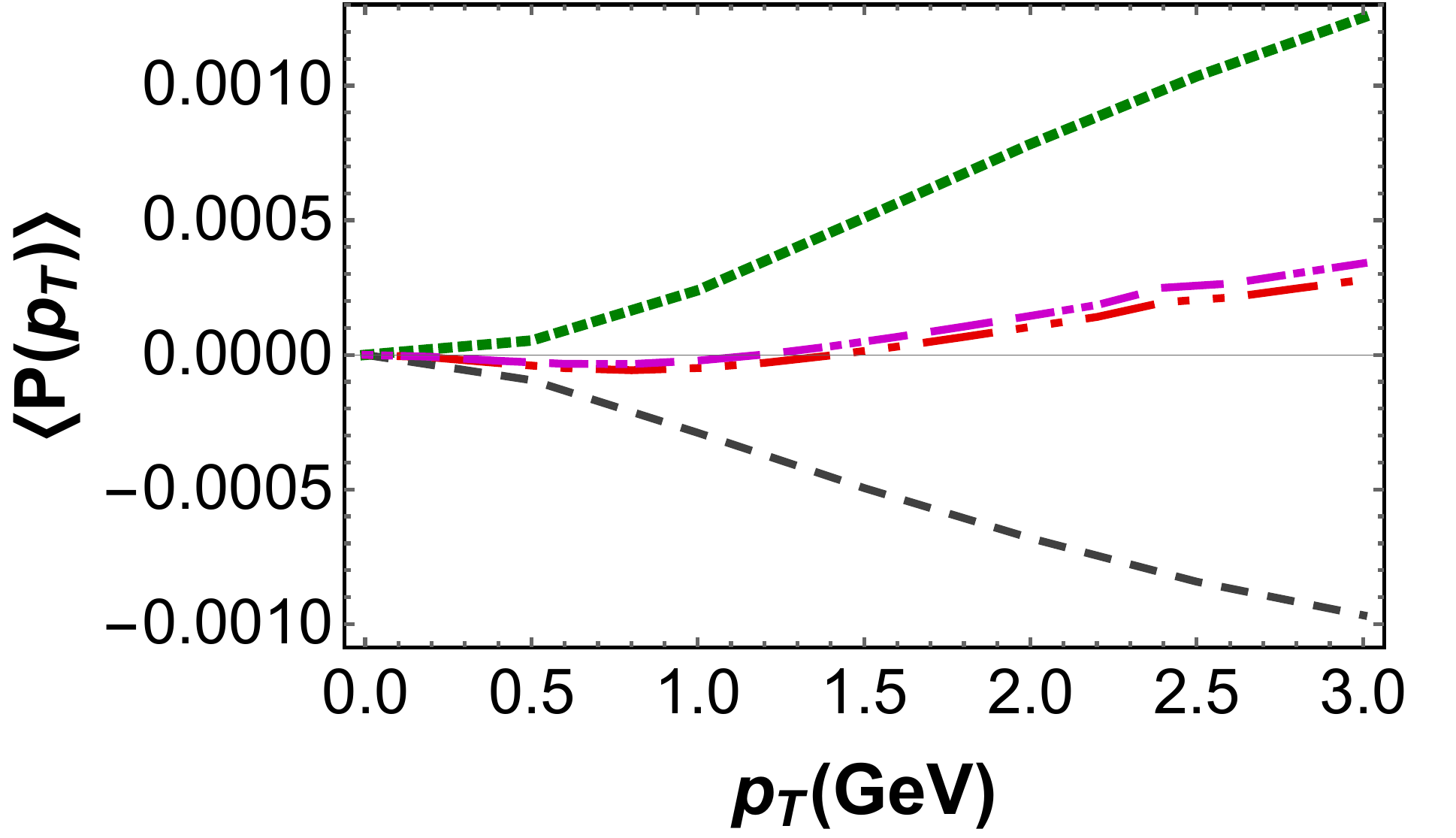}
		\caption{
		Same as Fig. \ref{fig:f2_1new} but for the centrality class  $c$=30--60\%. 
		(a) For comparison we show the dependence of longitudinal spin polarization of $\Lambda$ and $\bar{\Lambda}$ on azimuthal angle relative to second order event plane for the centrality class $c$=20--60\% plotted using the STAR data at $\sqrt{s_\text{NN}}=200\,\text{GeV}$~\cite{Adam:2019srw}.
		} 
	 \label{fig:f2_3new}
\end{figure*}

We have repeated calculations for $c$=15--30\% and $c$=30--60\% centrality bins in Figs.~\ref{fig:polarization2}, \ref{fig:f2_2new}, \ref{fig:polarization3}, and \ref{fig:f2_3new}. The overall trends are very similar across all centralities. 
To make a qualitative comparison with experimental data in Fig.~\ref{fig:f2_3new} we show
the longitudinal spin polarization of $\Lambda$ and $\bar{\Lambda}$ for the centrality class $c$=20--60\% measured by STAR experiment at $\sqrt{s_\text{NN}}=200\,\text{GeV}$~\cite{Adam:2019srw}. We see that the thermal shear component alone has the right sign, but a somewhat larger magnitude of polarization modulation than experimental data. However, when taken together with thermal vorticity contribution (with or without temperature gradients), the net results are too small to explain the observations.


In summary, we presented the study of recently introduced thermal shear contributions to longitudinal spin polarization of $\Lambda$ hyperons in a single freeze-out thermal model.
We confirm that the longitudinal polarization due to the thermal shear tensor alone has the same sign and similar magnitude to the one observed in STAR experiment. However, contributions of thermal vorticity and thermal shear are of opposite sign and nearly identically cancel each other if added together. We followed the prescription advocated in Ref.~\cite{Becattini:2021iol} and computed longitudinal polarization neglecting temperature gradients. Although we seem to get the same polarization sign as in the experimental data, the net predicted polarization is significantly below the data. We conclude that both thermal vorticity and thermal shear tensors make equally important contributions to longitudinal spin polarization and the net result depends sensitively to the procedure of how these two terms are combined.

\FloatBarrier
\begin{acknowledgments}
 AK acknowledge the post doctoral fellowship from the Indian Institute of Technology Gandhinagar, Gujarat, India. RR and WF were supported in part by the Polish National Science Centre Grants No. 2016/23/B/ST2/00717 and No. 2018/30/E/ST2/00432.
\end{acknowledgments}
\appendix
%%%%%%%%%%%%%%%%%%%%%%%%%%%%%%
\section{Expression for different components of $\xi^{II}_{\mu\nu}$ \label{sec:xitempgrad}}

 Here we list temperature gradients contribution to various components of thermal shear tensor.

\begin{widetext}
\beq
\xi^{II}_{01}&=&-\frac{tx}{2 T N^5} \Bigg(-\left(1+\delta -\sqrt{1+\delta}\right) \left(t^2-z^2+(1+\delta) x^2\right) -(1-\delta) \left(1+\delta +2 \sqrt{1+\delta }\right) y^2 \nn\\& &+2 c^2_s\Bigg(-\left(1+\delta -\sqrt{1+\delta}\right) \left(t^2-z^2+(1+\delta) x^2\right)+(1-\delta) \left(1+\delta+\sqrt{1+\delta }-2 \sqrt{1-\delta ^2}\right) y^2\Bigg)\Bigg) \nn\\
\eeq 
\beq
\xi^{II}_{02}&=&-\frac{t y}{2 T N^5}\Bigg(-\left(1-\delta -\sqrt{1-\delta }\right) \left(t^2-z^2+(1-\delta) y^2 \right)  +(1+\delta) \left(-2 \sqrt{1-\delta ^2}-\delta +\sqrt{1-\delta }+1\right) x^2\nn\\&&+2c^2_s\Bigg(\left(\sqrt{1-\delta ^2}-\delta +\sqrt{1-\delta }+1\right) \left(t^2-z^2\right) -\left(-\delta +2 \sqrt{1-\delta }+1\right) (\delta +1) x^2-(1-\delta )^{3/2} \left(\sqrt{\delta +1}+2\right) y^2\Bigg) \Bigg)\nn\\
\eeq
\beq
\xi^{II}_{03}&=&-\frac{tz}{2 T N^5}\Bigg( 2 \left(\sqrt{1-\delta }-1\right) (\delta -1) y^2-2 (\delta +1) \left(\sqrt{\delta +1}-1\right) x^2\nn\\&&+2c_s^2\Bigg(\left(\sqrt{1-\delta }+\sqrt{\delta +1}+1\right) \left(t^2-z^2\right) +(\delta -1) \left(\sqrt{\delta +1}+2\right) y^2-\left(\sqrt{1-\delta }+2\right) (\delta +1) x^2\Bigg) \Bigg)
\eeq
\beq
\xi^{II}_{12}&=&-\frac{x y}{2 T N^5}\Bigg(\left(-2 \sqrt{1-\delta ^2}+\delta  \left(\sqrt{1-\delta }-\sqrt{\delta +1}\right)+\sqrt{1-\delta }+\sqrt{\delta +1}\right) \left(t^2-z^2\right)\nn\\&+&(\delta +1)^{3/2} \left(\sqrt{1-\delta ^2}+\delta -1\right) x^2+(1-\delta )^{3/2} \left(\sqrt{1-\delta ^2}-\delta -1\right) y^2
\nn\\&&+2c_s^2\Bigg(\left(\sqrt{1-\delta ^2}+\delta  \left(\sqrt{1-\delta }-\sqrt{\delta +1}\right)+\sqrt{1-\delta }+\sqrt{\delta +1}\right) \left(z^2-t^2\right)\nn\\&+&\left(-\delta +2 \sqrt{1-\delta }+1\right) (\delta +1)^{3/2} x^2+(1-\delta )^{3/2} \left(\delta +2 \sqrt{\delta +1}+1\right) y^2\Big)\Bigg) \Bigg)
\eeq
\beq
\xi^{II}_{13}&=&-\frac{x z}{2 T N^5}\Bigg(\left(-\delta +\sqrt{\delta +1}-1\right) \left(-\left(t^2-z^2\right)\right)+(\delta -1) \left(-2 \sqrt{1-\delta ^2}+\delta +\sqrt{\delta +1}+1\right) y^2-(\delta +1) \left(-\delta +\sqrt{\delta +1}-1\right) x^2\nn\\&&+2c_s^2\Bigg(\left(\sqrt{1-\delta ^2}+\delta +\sqrt{\delta +1}+1\right) \left(-\left(t^2-z^2\right)\right) +\left(\sqrt{1-\delta }+2\right) (\delta +1)^{3/2} x^2-(\delta -1) \left(\delta +2 \sqrt{\delta +1}+1\right) y^2\Bigg)\Bigg)\nn\\
\eeq
\beq
\xi^{II}_{23}&=&\frac{1}{2 T N^5} \Bigg(\left(\delta +\sqrt{1-\delta }-1\right) \left(-\left(t^2-z^2\right)\right)+(\delta +1) \left(2 \sqrt{1-\delta ^2}+\delta -\sqrt{1-\delta }-1\right) x^2+(\delta -1) \left(\delta +\sqrt{1-\delta }-1\right) y^2\nn\\&+&2c_s^2\Bigg(\left(\sqrt{1-\delta ^2}-\delta +\sqrt{1-\delta }+1\right) \left(-\left(t^2-z^2\right)\right) +\left(-\delta +2 \sqrt{1-\delta }+1\right) (\delta +1) x^2+(1-\delta )^{3/2} \left(\sqrt{\delta +1}+2\right) y^2 \Bigg)\Bigg)\nn\\
\eeq
\beq
\xi^{II}_{00}&=&-\frac{t^2}{2 T N^5}\Bigg(2 (\delta +1) \left(\sqrt{\delta +1}-1\right) x^2-2 \left(\sqrt{1-\delta }-1\right) (\delta -1) y^2\nn\\&&+2c_s^2\Bigg(\left(\sqrt{1-\delta }+\sqrt{\delta +1}+1\right) \left(-\left(t^2-z^2\right)\right)+\left(\sqrt{1-\delta }+2\right) (\delta +1) x^2-(\delta -1) \left(\sqrt{\delta +1}+2\right) y^2\Bigg)\Bigg) 
\eeq
\beq
\xi^{II}_{11}&=&-\frac{x^2}{2 T N^5}\Bigg(2 (\delta +1) \left(\sqrt{\delta +1}-1\right) \left(t^2-z^2\right)+2 (\delta -1) \sqrt{\delta +1} \left(-\sqrt{1-\delta ^2}+\delta +1\right) y^2\nn\\&+&2c_s^2\Bigg((\delta +1) \left(\sqrt{1-\delta }+\sqrt{\delta +1}+1\right) \left(-\left(t^2-z^2\right)\right)+\left(\sqrt{1-\delta }+2\right) (\delta +1)^2 x^2-\left(\delta ^2-1\right) \left(\sqrt{\delta +1}+2\right) y^2\Bigg)\Bigg)\nn\\
\eeq
\beq
\xi^{II}_{22}&=&-\frac{y^2}{T N^5} \sqrt{1-\delta}\Bigg(\left(\delta +\sqrt{1-\delta }-1\right) (z^2-t^2)\nn\\&&-(\delta +1) \left(\sqrt{1-\delta ^2}+\delta -1\right) x^2+\sqrt{1-\delta }c_s^2\Bigg(\left(\sqrt{1-\delta }+1\right) z^2-2 (\delta -1) y^2+\sqrt{\delta +1} y^2\nn\\&&+\sqrt{\delta +1} z^2-\delta \sqrt{\delta +1} y^2+\left(\sqrt{1-\delta }+2\right) (\delta +1) x^2-\left(\sqrt{1-\delta }+\sqrt{\delta +1}+1\right) t^2\Bigg)\Bigg) \nn\\
\eeq
\beq
\xi^{II}_{33}&=&-\frac{z^2}{T N^5} \Bigg(\left(\sqrt{1-\delta }-1\right) (\delta -1) y^2+(\delta +1) \left(\sqrt{\delta +1}-1\right) x^2\nn\\&&+c_s^2\Bigg(\left(\sqrt{1-\delta }+1\right) z^2-2 (\delta -1) y^2+\sqrt{\delta +1} y^2+\sqrt{\delta +1} z^2-\delta  \sqrt{\delta +1} y^2\nn\\&&+\left(\sqrt{1-\delta }+2\right) (\delta +1) x^2-\left(\sqrt{1-\delta }+\sqrt{\delta +1}+1\right) t^2\Bigg) \Bigg)
\eeq
\end{widetext}
\bibliography{spin-lit}{}

\providecommand{\href}[2]{#2}\begingroup\raggedright\begin{thebibliography}{10}

\bibitem{Liang:2004xn}
Z.-T. Liang and X.-N. Wang, ``{Spin alignment of vector mesons in non-central
  A+A collisions},''
  \href{http://dx.doi.org/10.1016/j.physletb.2005.09.060}{{\em Phys. Lett.}
  {\bf B629} (2005)  20--26},
\href{http://arxiv.org/abs/nucl-th/0411101}{{\tt arXiv:nucl-th/0411101
  [nucl-th]}}.
%%CITATION = NUCL-TH/0411101;%%.

\bibitem{Voloshin:2004ha}
S.~A. Voloshin, ``{Polarized secondary particles in unpolarized high energy
  hadron-hadron collisions?},''
\href{http://arxiv.org/abs/nucl-th/0410089}{{\tt arXiv:nucl-th/0410089
  [nucl-th]}}.
%%CITATION = NUCL-TH/0410089;%%.

\bibitem{Voloshin:2017kqp}
S.~A. Voloshin, ``{Vorticity and particle polarization in heavy ion collisions
  (experimental perspective)},'' \href{http://arxiv.org/abs/1710.08934}{{\tt
  arXiv:1710.08934 [nucl-ex]}}.
[EPJ Web Conf.17,10700(2018)].
%%CITATION = ARXIV:1710.08934;%%.

\bibitem{Liang:2004ph}
Z.-T. Liang and X.-N. Wang, ``{Globally polarized quark-gluon plasma in
  non-central A+A collisions},''
  \href{http://dx.doi.org/10.1103/PhysRevLett.94.102301,
  10.1103/PhysRevLett.96.039901}{{\em Phys. Rev. Lett.} {\bf 94} (2005)
  102301}, \href{http://arxiv.org/abs/nucl-th/0410079}{{\tt
  arXiv:nucl-th/0410079 [nucl-th]}}.
[Erratum: Phys. Rev. Lett.96,039901(2006)].
%%CITATION = NUCL-TH/0410079;%%.

\bibitem{STAR:2017ckg}
{\bf STAR} Collaboration, L.~Adamczyk {\em et al.}, ``{Global $\Lambda$ hyperon
  polarization in nuclear collisions: evidence for the most vortical fluid},''
  \href{http://dx.doi.org/10.1038/nature23004}{{\em Nature} {\bf 548} (2017)
  62--65},
\href{http://arxiv.org/abs/1701.06657}{{\tt arXiv:1701.06657 [nucl-ex]}}.
%%CITATION = ARXIV:1701.06657;%%.

\bibitem{Adam:2018ivw}
{\bf STAR} Collaboration, J.~Adam {\em et al.}, ``{Global polarization of
  $\Lambda$ hyperons in Au+Au collisions at $\sqrt{s_{_{NN}}}$ = 200 GeV},''
  \href{http://dx.doi.org/10.1103/PhysRevC.98.014910}{{\em Phys. Rev.} {\bf
  C98} (2018)  014910},
\href{http://arxiv.org/abs/1805.04400}{{\tt arXiv:1805.04400 [nucl-ex]}}.
%%CITATION = ARXIV:1805.04400;%%.

\bibitem{Acharya:2019vpe}
{\bf ALICE} Collaboration, S.~Acharya {\em et al.}, ``{Measurement of
  spin-orbital angular momentum interactions in relativistic heavy-ion
  collisions},'' \href{http://dx.doi.org/10.1103/PhysRevLett.125.012301}{{\em
  Phys. Rev. Lett.} {\bf 125} (2020) no.~1, 012301},
  \href{http://arxiv.org/abs/1910.14408}{{\tt arXiv:1910.14408 [nucl-ex]}}.

\bibitem{Barnett:1935}
S.~J. Barnett, ``Gyromagnetic and electron-inertia effects,''
  \href{http://dx.doi.org/10.1103/RevModPhys.7.129}{{\em Rev. Mod. Phys.} {\bf
  7} (1935)  129--166}.
  \url{https://link.aps.org/doi/10.1103/RevModPhys.7.129}.

\bibitem{dehaas:1915}
A.~Einstein and W.~de~Haas, ``{Experimenteller Nachweis der Ampereschen
  Molekularstroeme},'' {\em Deutsche Physikalische Gesellschaft, Verhandlungen}
  {\bf 17} (1915)  152.

\bibitem{Becattini:2015ska}
F.~Becattini, G.~Inghirami, V.~Rolando, A.~Beraudo, L.~Del~Zanna, A.~De~Pace,
  M.~Nardi, G.~Pagliara, and V.~Chandra, ``{A study of vorticity formation in
  high energy nuclear collisions},''
  \href{http://dx.doi.org/10.1140/epjc/s10052-015-3624-1,
  10.1140/epjc/s10052-018-5810-4}{{\em Eur. Phys. J.} {\bf C75} (2015) no.~9,
  406}, \href{http://arxiv.org/abs/1501.04468}{{\tt arXiv:1501.04468
  [nucl-th]}}.
[Erratum: Eur. Phys. J.C78,no.5,354(2018)].
%%CITATION = ARXIV:1501.04468;%%.

\bibitem{Karpenko:2016jyx}
I.~Karpenko and F.~Becattini, ``{Study of $\Lambda $ polarization in
  relativistic nuclear collisions at $\sqrt{s_\mathrm {NN}}=7.7$ –200 GeV},''
  \href{http://dx.doi.org/10.1140/epjc/s10052-017-4765-1}{{\em Eur. Phys. J.}
  {\bf C77} (2017) no.~4, 213},
\href{http://arxiv.org/abs/1610.04717}{{\tt arXiv:1610.04717 [nucl-th]}}.
%%CITATION = ARXIV:1610.04717;%%.

\bibitem{Xie:2017upb}
Y.~Xie, D.~Wang, and L.~P. Csernai, ``{Global Lambda polarization in high
  energy collisions},''
  \href{http://dx.doi.org/10.1103/PhysRevC.95.031901}{{\em Phys. Rev.} {\bf
  C95} (2017) no.~3, 031901},
\href{http://arxiv.org/abs/1703.03770}{{\tt arXiv:1703.03770 [nucl-th]}}.
%%CITATION = ARXIV:1703.03770;%%.

\bibitem{Pang:2016igs}
L.-G. Pang, H.~Petersen, Q.~Wang, and X.-N. Wang, ``{Vortical Fluid and
  $\Lambda$ Spin Correlations in High-Energy Heavy-Ion Collisions},''
  \href{http://dx.doi.org/10.1103/PhysRevLett.117.192301}{{\em Phys. Rev.
  Lett.} {\bf 117} (2016) no.~19, 192301},
  \href{http://arxiv.org/abs/1605.04024}{{\tt arXiv:1605.04024 [hep-ph]}}.

\bibitem{Becattini:2017gcx}
F.~Becattini and I.~Karpenko, ``{Collective Longitudinal Polarization in
  Relativistic Heavy-Ion Collisions at Very High Energy},''
  \href{http://dx.doi.org/10.1103/PhysRevLett.120.012302}{{\em Phys. Rev.
  Lett.} {\bf 120} (2018) no.~1, 012302},
  \href{http://arxiv.org/abs/1707.07984}{{\tt arXiv:1707.07984 [nucl-th]}}.

\bibitem{Becattini:2016gvu}
F.~Becattini, I.~Karpenko, M.~Lisa, I.~Upsal, and S.~Voloshin, ``{Global
  hyperon polarization at local thermodynamic equilibrium with vorticity,
  magnetic field and feed-down},''
  \href{http://dx.doi.org/10.1103/PhysRevC.95.054902}{{\em Phys. Rev.} {\bf
  C95} (2017) no.~5, 054902},
\href{http://arxiv.org/abs/1610.02506}{{\tt arXiv:1610.02506 [nucl-th]}}.
%%CITATION = ARXIV:1610.02506;%%.

\bibitem{Adam:2019srw}
{\bf STAR} Collaboration, J.~Adam {\em et al.}, ``Polarization of $\lambda$
  ($\bar{\Lambda}$) hyperons along the beam direction in au+au collisions at
  $\sqrt{s{_{NN}}}$ = 200 gev,''
  \href{http://dx.doi.org/10.1103/PhysRevLett.123.132301}{{\em Phys.Rev.Lett.}
  {\bf 123} (2019) no.~13, 132301}, \href{http://arxiv.org/abs/1905.11917}{{\tt
  arXiv:1905.11917 [nucl-ex]}}.

\bibitem{Becattini:2020ngo}
F.~Becattini and M.~A. Lisa, ``{Polarization and Vorticity in the
  Quark\textendash{}Gluon Plasma},''
  \href{http://dx.doi.org/10.1146/annurev-nucl-021920-095245}{{\em Ann. Rev.
  Nucl. Part. Sci.} {\bf 70} (2020)  395--423},
  \href{http://arxiv.org/abs/2003.03640}{{\tt arXiv:2003.03640 [nucl-ex]}}.

\bibitem{Niida:2018hfw}
{\bf STAR} Collaboration, T.~Niida,
  \href{http://dx.doi.org/10.1016/j.nuclphysa.2018.08.034}{``Global and local
  polarization of $\lambda$ hyperons in au+au collisions at 200 gev from
  star,''} in {\em Global and local polarization of $\Lambda$ hyperons in Au+Au
  collisions at 200 GeV from STAR}, vol.~982, pp.~511--514.
\newblock 2019.
\newblock \href{http://arxiv.org/abs/1808.10482}{{\tt arXiv:1808.10482
  [nucl-ex]}}.

\bibitem{Weickgenannt:2020aaf}
N.~Weickgenannt, E.~Speranza, X.-l. Sheng, Q.~Wang, and D.~H. Rischke,
  ``{Generating spin polarization from vorticity through nonlocal
  collisions},'' \href{http://arxiv.org/abs/2005.01506}{{\tt arXiv:2005.01506
  [hep-ph]}}.

\bibitem{Speranza:2020ilk}
E.~Speranza and N.~Weickgenannt, ``{Spin tensor and pseudo-gauges: from nuclear
  collisions to gravitational physics},''
  \href{http://arxiv.org/abs/2007.00138}{{\tt arXiv:2007.00138 [nucl-th]}}.

\bibitem{Weickgenannt:2021cuo}
N.~Weickgenannt, E.~Speranza, X.-l. Sheng, Q.~Wang, and D.~H. Rischke,
  ``{Derivation of the nonlocal collision term in the relativistic Boltzmann
  equation for massive spin-1/2 particles from quantum field theory},''
  \href{http://arxiv.org/abs/2103.04896}{{\tt arXiv:2103.04896 [nucl-th]}}.

\bibitem{Becattini:2021iol}
F.~Becattini, M.~Buzzegoli, A.~Palermo, G.~Inghirami, and I.~Karpenko, ``{Local
  polarization and isothermal local equilibrium in relativistic heavy ion
  collisions},'' \href{http://arxiv.org/abs/2103.14621}{{\tt arXiv:2103.14621
  [nucl-th]}}.

\bibitem{Fu:2021pok}
B.~Fu, S.~Y.~F. Liu, L.~Pang, H.~Song, and Y.~Yin, ``{Shear-Induced Spin
  Polarization in Heavy-Ion Collisions},''
  \href{http://dx.doi.org/10.1103/PhysRevLett.127.142301}{{\em Phys. Rev.
  Lett.} {\bf 127} (2021) no.~14, 142301},
  \href{http://arxiv.org/abs/2103.10403}{{\tt arXiv:2103.10403 [hep-ph]}}.

\bibitem{Becattini:2021suc}
F.~Becattini, M.~Buzzegoli, and A.~Palermo, ``{Spin-thermal shear coupling in a
  relativistic fluid},''
  \href{http://dx.doi.org/10.1016/j.physletb.2021.136519}{{\em Phys. Lett. B}
  {\bf 820} (2021)  136519}, \href{http://arxiv.org/abs/2103.10917}{{\tt
  arXiv:2103.10917 [nucl-th]}}.

\bibitem{Liu:2021uhn}
S.~Y.~F. Liu and Y.~Yin, ``{Spin polarization induced by the hydrodynamic
  gradients},'' \href{http://dx.doi.org/10.1007/JHEP07(2021)188}{{\em JHEP}
  {\bf 07} (2021)  188}, \href{http://arxiv.org/abs/2103.09200}{{\tt
  arXiv:2103.09200 [hep-ph]}}.

\bibitem{Broniowski:2001we}
W.~Broniowski and W.~Florkowski, ``{Explanation of the RHIC p(T) spectra in a
  thermal model with expansion},''
  \href{http://dx.doi.org/10.1103/PhysRevLett.87.272302}{{\em Phys. Rev. Lett.}
  {\bf 87} (2001)  272302}, \href{http://arxiv.org/abs/nucl-th/0106050}{{\tt
  arXiv:nucl-th/0106050}}.

\bibitem{Cleymans:1992zc}
J.~Cleymans and H.~Satz, ``{Thermal hadron production in high-energy heavy ion
  collisions},'' \href{http://dx.doi.org/10.1007/BF01555746}{{\em Z. Phys. C}
  {\bf 57} (1993)  135--148}, \href{http://arxiv.org/abs/hep-ph/9207204}{{\tt
  arXiv:hep-ph/9207204}}.

\bibitem{Braun-Munzinger:2001hwo}
P.~Braun-Munzinger, D.~Magestro, K.~Redlich, and J.~Stachel, ``{Hadron
  production in Au - Au collisions at RHIC},''
  \href{http://dx.doi.org/10.1016/S0370-2693(01)01069-3}{{\em Phys. Lett. B}
  {\bf 518} (2001)  41--46}, \href{http://arxiv.org/abs/hep-ph/0105229}{{\tt
  arXiv:hep-ph/0105229}}.

\bibitem{Florkowski:2001fp}
W.~Florkowski, W.~Broniowski, and M.~Michalec, ``{Thermal analysis of particle
  ratios and p(t) spectra at RHIC},'' {\em Acta Phys. Polon. B} {\bf 33} (2002)
   761--769, \href{http://arxiv.org/abs/nucl-th/0106009}{{\tt
  arXiv:nucl-th/0106009}}.

\bibitem{Becattini:2005xt}
F.~Becattini, J.~Manninen, and M.~Gazdzicki, ``{Energy and system size
  dependence of chemical freeze-out in relativistic nuclear collisions},''
  \href{http://dx.doi.org/10.1103/PhysRevC.73.044905}{{\em Phys. Rev. C} {\bf
  73} (2006)  044905}, \href{http://arxiv.org/abs/hep-ph/0511092}{{\tt
  arXiv:hep-ph/0511092}}.

\bibitem{Andronic:2017pug}
A.~Andronic, P.~Braun-Munzinger, K.~Redlich, and J.~Stachel, ``{Decoding the
  phase structure of QCD via particle production at high energy},''
  \href{http://dx.doi.org/10.1038/s41586-018-0491-6}{{\em Nature} {\bf 561}
  (2018) no.~7723, 321--330}, \href{http://arxiv.org/abs/1710.09425}{{\tt
  arXiv:1710.09425 [nucl-th]}}.

\bibitem{Florkowski:2019voj}
W.~Florkowski, A.~Kumar, R.~Ryblewski, and A.~Mazeliauskas, ``{Longitudinal
  spin polarization in a thermal model},''
  \href{http://dx.doi.org/10.1103/PhysRevC.100.054907}{{\em Phys. Rev. C} {\bf
  100} (2019) no.~5, 054907}, \href{http://arxiv.org/abs/1904.00002}{{\tt
  arXiv:1904.00002 [nucl-th]}}.

\bibitem{Voloshin}
S.~Voloshin, ``{International Workshop XLVII on Gross Properties of Nuclei and
  Nuclear Excitations},''
\newblock Hirschegg, Kleinwalsertal, Austria, January 13-19, 2019.

\bibitem{Broniowski:2002wp}
W.~Broniowski, A.~Baran, and W.~Florkowski, ``{Thermal model at RHIC. Part 2.
  Elliptic flow and HBT radii},''
  \href{http://dx.doi.org/10.1063/1.1570571}{{\em AIP Conf. Proc.} {\bf 660}
  (2003) no.~1, 185--195},
\href{http://arxiv.org/abs/nucl-th/0212053}{{\tt arXiv:nucl-th/0212053
  [nucl-th]}}.
%%CITATION = NUCL-TH/0212053;%%.

\bibitem{baran}
A.~Baran, ``{Description of azimuthal asymmetry in relativistic heavy-ion
  collisions based on a thermal model of particle production},''
\newblock PhD Thesis (W. Broniowski - supervisor), Institute of Nuclear
  Physics, Krakow, Poland, 2004.

\bibitem{Florkowski:2004du}
W.~Florkowski, W.~Broniowski, and A.~Baran, ``{Strange particle production in a
  single-freeze-out model},''
  \href{http://dx.doi.org/10.1088/0954-3899/31/6/064}{{\em J. Phys.} {\bf G31}
  (2005)  S1087--S1090},
\href{http://arxiv.org/abs/nucl-th/0412077}{{\tt arXiv:nucl-th/0412077
  [nucl-th]}}.
%%CITATION = NUCL-TH/0412077;%%.

\bibitem{Florkowski:2018ahw}
W.~Florkowski, A.~Kumar, and R.~Ryblewski, ``{Thermodynamic versus kinetic
  approach to polarization-vorticity coupling},''
  \href{http://dx.doi.org/10.1103/PhysRevC.98.044906}{{\em Phys. Rev.} {\bf
  C98} (2018)  044906},
\href{http://arxiv.org/abs/1806.02616}{{\tt arXiv:1806.02616 [hep-ph]}}.
%%CITATION = ARXIV:1806.02616;%%.

\end{thebibliography}\endgroup
\bibliographystyle{utphys}

\end{document}